\documentclass[]{interact}

\usepackage{epstopdf}

\usepackage[numbers,sort&compress]{natbib}
\bibpunct[, ]{[}{]}{,}{n}{,}{,}
\renewcommand\bibfont{\fontsize{10}{12}\selectfont}
\makeatletter
\def\NAT@def@citea{\def\@citea{\NAT@separator}}
\makeatother

\theoremstyle{plain}

\theoremstyle{definition}

\theoremstyle{remark}
\newtheorem{remark}{Remark}

\usepackage{multirow}
\usepackage[symbol]{footmisc}
\usepackage{amsmath}
\usepackage{graphicx}      
\usepackage{algpseudocode,algorithm,algorithmicx}

\algrenewcommand\algorithmicrequire{\textbf{Precondition:}}
\algrenewcommand\algorithmicensure{\textbf{Postcondition:}}
\usepackage[caption=false,font=footnotesize]{subfig}

\usepackage{tikz}
\usepackage{amssymb}  

\usetikzlibrary{shapes,positioning}
\usetikzlibrary{narrow}
\usetikzlibrary{shapes.misc}
\usepackage{booktabs}{\tiny }
\usepackage{pgfplots} 
\usepackage{reglertikz}
\usepackage{pgf}
\usepackage{mathsymbols}
\usetikzlibrary{external}
\newcommand\figdir{./figs}
\tikzset{external/figure list=true}
\tikzset{external/up to date check=simple} 
\usepackage{filemod}
\pgfplotsset{compat=newest}
\newcommand{\tikzcustomremake}[2]{%
	\tikzset{external/remake next}%
}
\makeatletter
\renewcommand*\env@matrix[1][*\c@MaxMatrixCols c]{%
	\hskip -\arraycolsep
	\let\@ifnextchar\new@ifnextchar
	\array{#1}}
\makeatother

\newif\ifconfidential

\confidentialfalse

\ifconfidential
\usepackage{fancyhdr}
\fi

\newcommand{\red}[1]{\textcolor{black}{#1}}

\definecolor{wheat}{rgb}{0.96,0.87,0.70}

\newcommand{\software}[1]{\texttt{#1}}

\begin{document}
	
\articletype{ARTICLE DRAFT}

\title{Friction-Adaptive Stochastic Nonlinear Model Predictive Control for Autonomous Vehicles}
\author{
	\name{Sean Vaskov\textsuperscript{a,b}\thanks{CONTACT Karl Berntorp. Email: karl.o.berntorp@ieee.org}, , Rien Quirynen\textsuperscript{a}, \red{Marcel Menner\textsuperscript{a}}, and Karl Berntorp\textsuperscript{a}}
	\affil{\textsuperscript{a}Mitsubishi Electric Research Laboratories, Cambridge, MA, 02139, USA.; \textsuperscript{b}University of Michigan, Ann Arbor, MI, 48109, USA.}
}
%
%
\ifconfidential
\fancyhead[C]{{\bf MERL CONFIDENTIAL}. $^\copyright$MERL, $2021$. }
\fancyfoot[C]{{\bf MERL CONFIDENTIAL}. $^\copyright$MERL, $2021$}
\fi

\ifconfidential
\thispagestyle{fancy}
\pagestyle{fancy}
\else
\thispagestyle{empty}
\pagestyle{empty}
\fi

\maketitle

\begin{abstract}
This paper addresses the trajectory-tracking problem under uncertain road-surface conditions for autonomous vehicles. We propose a stochastic nonlinear model predictive controller (SNMPC) that learns a tire--road friction model online using standard automotive-grade sensors. Learning the entire tire--road friction model in real time requires driving in the nonlinear, potentially unstable regime of the vehicle dynamics, using a prediction model that may not have fully converged. To handle this, we formulate the tire-friction model learning in a Bayesian framework, and propose two estimators that learn different aspects of the tire--road friction. The estimators output the estimate of the tire-friction model as well as the uncertainty of the estimate, which expresses the confidence in the model for different driving regimes. The SNMPC exploits the uncertainty estimate in its prediction model to take proper action when the uncertainty is large. We validate the approach in an extensive Monte-Carlo study using real vehicle parameters \red{and in CarSim}. The results when comparing to various MPC approaches indicate a substantial reduction in constraint violations, as well as a reduction in closed-loop cost. We also demonstrate the real-time feasibility in automotive-grade processors using a dSPACE MicroAutoBox-II rapid prototyping unit, showing a worst-case computation time of roughly 40ms.
\end{abstract}

\section{INTRODUCTION}
\label{sec:Intro}
The strong push recently in the automotive industry for introducing new technologies related to automated driving (AD) and advanced driving-assistance systems (ADAS), has led to predictive information being readily available to the vehicle controllers. For instance, onboard sensors detect the environment in the vicinity of the vehicle, 
motion planners generate future desired driving profiles \cite{Paden2016,Berntorp2017c}, and motion-prediction modules provides risk assessments for different control actions \cite{Lefevre2014,Okamoto2017a}. Consequently, model predictive control (MPC) is suitable for vehicle control as it naturally integrates predictive information in its problem formulation \cite{Dicairano2018,Hrovat2012}, and MPC has successfully been applied to vehicle trajectory tracking and stability control \cite{Falcone2007,Berntorp2019f,Carvalho2015,DiCairano2016,Borrelli2005}.

The key benefit of MPC for vehicle-control applications is its ability to naturally incorporate constraint handling as well as prediction models in its optimal control problem~(OCP). However, because MPC relies on a vehicle model, it is important that the model represents sufficiently well the actual vehicle. Specifically, since the main actuation of the vehicle is done by altering the forces generated by the tire--road contact, it is imperative to have a sensible guess about the road surface the car operates on, as otherwise safe operation of the vehicle can be endangered \cite{Berntorp2014,Quirynen2018a}. 

\begin{figure}
	\includegraphics{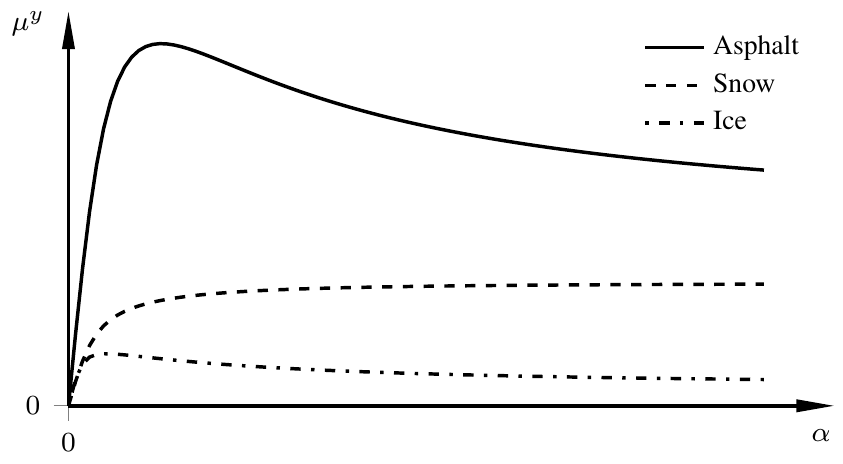}
	\centering
	\includegraphics{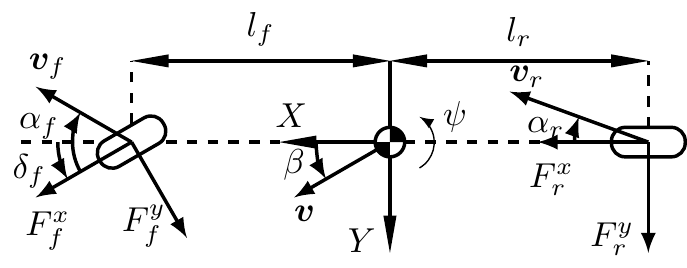}
	\caption{ {\bf Upper:} Examples of lateral friction as a function of 
		slip angle $\alpha$ for asphalt, loose snow, and ice. {\bf Lower:} A schematic of the single-track 
		vehicle model and related notation.}\label{fig:pacejka}\label{fig:ST}
\end{figure}

In this paper, we develop an MPC that adapts in real time to different road-surface conditions. The interaction between tire and road is highly nonlinear, and the function describing the nonlinear relation varies heavily based on the road surface and other tire properties~\cite{Gustafsson1997,Svendenius2007}. Fig.~\ref{fig:pacejka} shows examples of the function relating the lateral tire force to the tire side slip angle, for different surfaces. The force-slip relation is approximately linear for small slip values, which are typical when driving in normal conditions.
However, when driving close to the adhesion limits, which may happen in emergency maneuvers, on unpaved road, or on wet and icy roads, the nonlinear part of the tire-force function may be excited, and hence the full tire curve shape must be considered. Thus, when driving over different surfaces the time-varying tire-force curve needs to be identified in real time, rapidly, and using noisy onboard sensing.

A complicating factor in identifying nonlinear functions is that the entire range of the function should be excited to generate data for identification. In case of the tire force function, obtaining data for the nonlinear part is challenging as it requires to drive the vehicle to the limits of its performance envelope and, usually, this is not done unless an emergency maneuver is needed or a particularly challenging surface is encountered. Furthermore, driving in the nonlinear region of the tire force function before a reliable model of the same is obtained is challenging and possibly dangerous. If the controller approaches the nonlinear region of the force curve without having a reliable model, this can cause control errors possibly leading to vehicle instability. In addition, a difficulty when learning the tire-friction function using automotive-grade sensors is that the amount of sensors is limited, and they are relatively low grade \cite{Gustafsson2009}. Moreover, not only do the sensors only provide indirect measurements of the friction, they do not even measure the vehicle state, which is nonlinearly dependent on the tire friction and must therefore be known for learning the tire friction. 

To address these issues, in this paper, we couple online tire-friction estimation with a stochastic nonlinear MPC (SNMPC). The proposed SNMPC models the tire-road friction as a disturbance affecting the prediction model and is flexible in the sense that as long as a tire-friction estimator can output the tire-road friction model by the first two moments (i.e., mean and covariance), the SNMPC can readily incorporate the estimates. As an example of this, we integrate the SNMPC with two Bayesian estimators capable of outputting the first two moments; (i), a computationally efficient tire-stiffness estimator modeling the tire-road friction as a linear function of the slip; and (ii), a Gaussian-process (GP) based estimator with unknown and time-varying mean and covariance function \cite{Rasmussen2006}, leading to a GP state-space model (GP-SSM). In particular, we leverage two recently proposed methods for real-time joint state estimation and learning of the state-transition function in dynamical systems \cite{Berntorp2018h,Berntorp2021d} based on particle filtering \cite{Doucet2009} for jointly estimating online the state and associated tire-friction model. \red{These two methods are similar in the sense that they estimate the distribution and associated weighted mean of the tire friction using only sensors available in production cars. Since the methods are based on particle filtering, they have asymptotic convergence guarantees.} Due to the estimators being Bayesian, they behave in a manner suitable for safety-critical vehicle control, as they predict larger uncertainty for parts of the system that have not been sufficiently excited. 

The resulting control system leverages the information from the estimator in its prediction model. The respective estimator learns the tire-friction function using a model-focused approach where it updates the tire-friction model as data are gathered, which leads to efficient learning where only a few data points are needed to achieve reliable estimates, resulting in an efficient and safe closed-loop behavior suitable for real-time control. The SNMPC is based on an efficient block-sparse quadratic programming (QP) solver \cite{Quirynen2020} for use within nonlinear optimal control, and the SNMPC with probabilistic chance constraints uses a tailored Jacobian approximation along with an adjoint-based sequential QP (SQP) optimization method \cite{Feng2020}. \red{The enabler for integrating the friction estimators with the SNMPC formulation is the insight that although the filters estimate different aspects of the tire-friction relation, they can both output the mean and covariance of the respective friction quantitites. This allows a simple and intutive connection to the SNMPC formulation.} 

We have previously reported on methods for friction-adaptive MPC in~\cite{Berntorp2019f,Vaskov2021,Vaskov2022}. The current paper differs in that the work in~\cite{Berntorp2019f} uses tire-stiffness estimates to choose from a fixed library of predetermined tire-friction functions and does not utilize SNMPC. In \cite{Vaskov2021}, we developed a friction estimator targeting the linear region of the tire-friction curve~\cite{Berntorp2018h}, and in \cite{Vaskov2022} we showed how SNMPC can be coupled with GP-based particle filtering for controlling the vehicle while estimating the full tire-friction curve. This paper extends our preliminary work \cite{Vaskov2021,Vaskov2022} in three aspects. 
\begin{itemize}
	\item We generalize the estimation formulation and show that the SNMPC can be connected to any tire-friction estimator, as long as such estimator can output a mean estimate and associated covariance of the tire-friction model.
	\item We provide an extensive comparison of the closed-loop performance using two different estimators and discuss pros and cons for each of the methods.
	\item We show that the proposed control system is real-time feasible for current automotive micro-controllers, by implementing it on a dSPACE MicroAutoBox-II rapid prototyping unit, where the worst-case computation time for the proposed vehicle control system is about 40ms, irrespective of the estimator being used.
\end{itemize}
\subsection{Outline}
We outline the estimation and control model, together with the problem formulation in Sec.~\ref{sec:modeling}. Sec.~\ref{sec:BL} details an overview of the tire-friction estimators employed in the paper. Sec.~\ref{sec:SNMPC} gives the SNMPC problem formulation and implementation details based on the control model, and we make the connection between the SNMPC and the tire-friction estimator and outline the control system in Sec.~\ref{sec:RCS}, which is evaluated in a Monte-Carlo study in Sec.~\ref{sec:res}. Finally, the paper concludes with Sec.~\ref{sec:conclusion}.
\subsection{Notation}
The notation $\mathcal{R}(a)$ means the 2D rotation matrix of an angle 
$a$. Vectors are shown in bold, $\xvec$, we denote the stacking of two vectors $\avec$, 
$\bvec$ by $[\avec,\bvec]$, and constraints between vectors are intended componentwise. Throughout, $\xvec \sim \mathcal{N}(\hat \xvec,\Sigmabf)$ 
indicates that $\xvec\in \Real^{n_x}$ is Gaussian distributed with mean $\hat \xvec$ and covariance $\mathrm{cov}(\xvec)=\Sigmabf$. Matrices are indicated in capital bold font as $\X$. 
With $p(\xvec_{0:k}|\yvec_{0:k})$, we mean the posterior density function of the state trajectory $\xvec_{0:k}$ from time step 0 to time step $k$ given the measurement sequence ${\yvec_{0:k} :=\{\yvec_0,\ldots,\yvec_k\}}$, and $\xvec_{0:k}^i$ is the $i^{\text{th}}$ realization of $\xvec_{0:k}$.
The notation $\fvec(\xvec) \sim \mathcal{GP}(\hat \fvec(\xvec),\Sigmabf(\xvec))$ means that the function $\fvec(\xvec)$ is a realization from a GP with a mean function $\hat \fvec(\cdot)$ and covariance $\Sigmabf(\cdot)$. For a continuous-time signal $\xvec(t)$ 
sampled with period $T_s$, $x_k$ denotes the $k^{\rm th}$ sample, i.e., $\xvec_k=\xvec(kT_s)$ and 
$\xvec_{k+h|k}$ is 
the value of $\xvec$ predicted $h$ steps ahead from $k$, i.e., the predicted value of 
$\xvec((k+h)T_s)$ 
based on $ \xvec(kT_s)$.
The quadratic form of the squared weighted norm of a vector $\xvec$ and matrix $Q$ is given by $\|\xvec\|_Q^2 = \xvec^\top Q\xvec$.

\section{Modeling and Problem Formulation}
\label{sec:modeling}
The vehicle model is composed of a chassis model 
describing the motion of the rigid body due to the forces generated at the tires and a tire model describing 
what forces the tires generate depending on the chassis and wheel velocities.
For the chassis, we consider the standard single-track model, where the left and 
right track of the car are lumped into a single centered track, as Fig.~\ref{fig:ST} shows. Hence, only 
a single front and a single rear tire are considered, and roll and pitch dynamics are ignored, resulting 
in two translational and one rotational degrees of freedom. While for performance driving it 
may be advantageous to use a double-track chassis model, which includes lateral and longitudinal 
load transfer~\cite{Berntorp2014}, in~\cite{Berntorp2014,Quirynen2018a} the single-track model was 
shown to be sufficiently accurate for regular driving conditions, including when
tire forces are in the nonlinear region, because in such conditions the roll and pitch 
angles 	remain relatively small. Similarly, the single-track model seems sufficient in most evasive maneuvers, because the focus of such maneuvers is on preserving safety rather than achieving optimality, and hence a high-precision model is often unnecessary. On 
the other hand, the single-track model results in a 
reduced computing load, which is always desirable in 
automotive applications~\cite{Dicairano2018}, particularly for evasive maneuvers.

Taking the longitudinal and lateral velocities in the vehicle frame, $v^X$,
$v^Y$, and the yaw rate, $\dot \psi$, as states, the single-track model is described 
by
\begin{subequations} 
	\begin{align}
		\dot{v}^X - v^Y \dot{\psi}  &= \frac{1}{m} ( F_{f}^x\cos(\delta_f) +
		F_{r}^x -F_{f}^y\sin(\delta_f) ) ,  \label{eq:st1a}\\
		\dot{v}^Y + v^X \dot{\psi}  &= \frac{1}{m} ( F_{f}^y\cos(\delta_f) +
		F_{r}^y +F_{f}^x\sin(\delta_f) )  , \label{eq:st1b}\\
		I_{zz} \ddot{\psi}  &= l_f F_{f}^y\cos(\delta_f) - l_r F_{r}^y +
		l_fF_{f}^x\sin(\delta_f),\label{eq:st1c}
	\end{align}\label{eq:st1}
\end{subequations}
where $F^x_i=F_i^z\mu^x_i(\lambda_i)$, $F^y_i=F_i^z\mu^y_i(\alpha_i)$ are the total longitudinal/lateral forces in the tire frame for the lumped left and 
right tires where the friction functions $\mu^{x}_i$, $\mu^{y}_i$ are provided by the estimator, $\lambda_i$ is the longitudinal wheel slip, $\alpha_i$ is the slip angle, and subscripts $i=f,r$ denote front and rear, respectively. The 
normal forces resting on the lumped front/rear wheels $F_i^z$, $i\in\{f,r\}$, are
$F_{f}^z =  mg{l_r}/{l}$, $F_{r}^z =  mg{l_f}/{l}$, 
where $g$ is the gravity acceleration, $l_f$, $l_r$ are the distances of 
front and rear axles from the center of gravity, and $l = l_f + l_r$ is the vehicle wheel base. Also, $m$ is the vehicle mass, 
$I_{zz}$ is the vehicle inertia about the vertical axis, and $\delta_f$ is the front wheel (road) steering 
angle. The vehicle position in global coordinates $\pvec= (p^{\bf X}, p^{\bf Y})$ is obtained from the 
kinematic equation
\begin{equation}
	\begin{bmatrix} \dot p^{\bf X}\\ \dot p^{\bf Y} \end{bmatrix}= 
	\mathcal{R}(\psi) 
	\begin{bmatrix} v^X\\ v^Y \end{bmatrix}.
\end{equation}
The slip angles~$\alpha_{i}$ and slip ratios~$\lambda_{i}$ are
defined as in~\cite{Rajamani2006},
\begin{align}
	\alpha_i= -\arctan \left( \frac{v_i^y}{v_i^x} \right),\quad \lambda_i = \frac{R_w \omega_i - v_i^x}{\max(R_w\omega_i,v_i^x)},\label{eq:alphalambda} 
\end{align}
where $R_w$ is the wheel radius, and $v_i^x$ and $v_i^y$ are the longitudinal and lateral wheel velocities for wheel $i$. Given the velocity vector at the center of mass, $v = [v^X\ 
v^Y]^\top$, the velocity vectors at the wheels are 
\begin{equation}\label{eq:vwheel}
\begin{bmatrix} v^x_{i}\\ v^y_{i} \end{bmatrix}= \mathcal{R}(\delta_i)^\top 
		\begin{bmatrix} v^X\\ v^Y+c_i\dot \psi \end{bmatrix},\  c_f=l_f,\ c_r=- l_r. 
\end{equation}
\subsection{Problem Formulation}
We consider a scenario where we  
measure the vehicle position $(p^{\bf X}, p^{\bf Y})$ and the yaw (heading) $\psi$ (e.g., by GPS), the 
individual wheel speeds $\omega_{i,j}$, $i\in\{f,r\}$, where $j\in\{l,r\}$ denotes left and 
right tires, by wheel encoders, which are used to determine also the vehicle 
speed, approximately equal to $v^X$. We
measure the vehicle
longitudinal, $a^X=\dot{v}^X - v^Y \dot{\psi} $, and lateral, $a^Y=\dot{v}^Y + v^X \dot{\psi} $, 
acceleration and the yaw rate $\dot \psi$ by an automotive-grade inertial measurement unit
and the front road wheel steering angle $\delta_f$ by a relative encoder.
The control inputs are the front and rear wheel speeds~$\omega_f,\ \omega_r$ and the tire-wheel angle rate of change~\red{$\dot \delta_f$}.

We target for implementation of our proposed control system on automotive micro-controllers, which, due to harsh operating
conditions, hard real-time requirements, and cost considerations, are significantly less 
capable than desktop 	
computers~\cite{Dicairano2018} in terms of memory and speed. Hence, we aim at limiting
the amount of computations and data storage needed by our approach.

Under the above conditions, the objective of this paper is to design a control strategy 
that makes the vehicle motion follow 
a time-dependent reference trajectory $(p^{\bf X}_\mathrm{ref}(\cdot), p^{\bf Y}_\mathrm{ref}(\cdot), 
\psi_\mathrm{ref}(\cdot), v^X_\mathrm{ref}(\cdot))$, possibly generated in real 
time with an adequate preview, while operating over different road surfaces and 
environmental conditions. 
Since some of the vehicle states, e.g., $v^Y$, are not directly measured and the tire-friction model
is unknown (or at the very least uncertain)
and changes over time, the control strategy needs to 
estimate the vehicle state and the tire-friction model in real time with the controller activated.

\section{Bayesian Tire-Friction Learning}
\label{sec:BL}
In this section, we discuss the estimation framework used in conjunction with the SNMPC to formulate our control strategy. Our SNMPC formulation assumes that the uncertainty of the tire-friction model estimate can be well represented by a mean estimate and associated covariance. We present two methods for real-time estimation of a tire-friction model that both satisfy the requirements of the SNMPC. Both estimators are targeted for embedded automotive-grade hardware and sensors. In fact, they use the same sensor setup and the only difference being the complexity of the tire-friction model it estimates. A difficulty when addressing the tire-friction estimation problem using automotive-grade sensors is that the amount of sensors is limited, and they are relatively low grade \cite{Gustafsson2009}. Moreover, not only do the sensors only provide indirect measurements of the friction, they do not even measure the vehicle state, which is nonlinearly dependent on the tire friction and must therefore be known for learning the tire friction. Subsequently, we employ a recently developed method for \emph{jointly} estimating the tire-friction function $\mubf$ and the vehicle state $\xvec$ only using sensors available in production cars, namely wheel-speed sensors and inexpensive accelerometers and gyroscopes. 
We briefly outline the formulation of the respective method and refer to \cite{Berntorp2021d,Berntorp2020a,Berntorp2018h} for a more complete description.


\begin{remark}
	In this paper, we focus on the estimation of the lateral tire friction at front and rear wheels (i.e., $n_\mathrm{\mu} = 2$). The extension to the longitudinal case is analogous. \red{Hence, in the estimation, we }focus on the lateral vehicle dynamics, because usually these are the most critical and challenging for vehicle stability and ADAS.
\end{remark}
\subsection{Estimation Model}
The estimation model is based on the single-track equations in~\eqref{eq:st1} and models the tire-friction components as static functions of the slip quantities, which in the lateral case results in 
\begin{equation}\label{eq:forcedef}
	\mu^y_{i} = f_i^y(\alpha_i(\xest,\uest)), \quad i \in \{f,r\},
\end{equation}
where the state and input vector for the estimator are defined as $\xest=[v^Y,\dot \psi]^\top$, $\uest=[\delta,v^X]^\top$. For brevity, we define the vector $\alphabf = [\alpha_f ,\alpha_r]^\top$.

The two estimators we employ only differ in the way they model \eqref{eq:forcedef}. In the first estimator, the tire-friction model is linear, such that the friction reads as
\begin{equation}\label{eq:stiffness}
	\mu^y_{i} = C^y_{i} \alpha_i, i \in \{f,r\}.
\end{equation}
The stiffness vector {$\C = [C^y_{f}, C^y_{r}]^\top$} in \eqref{eq:stiffness} is for each component decomposed into a nominal and unknown part, \begin{equation}
	C_i^y = C_{i,n}^y+\Delta C_{i,k},
	\label{eq:stiffness2}
\end{equation}
where $C_{i,n}^y$ is the nominal value of the tire stiffness, for example, a priori determined on a nominal surface, and $\Delta C_{i,k}$ is a time-varying, unknown part,
which for each time step $k$ is Gaussian distributed according to $\Delta \C \sim \Ncal(\hat{\C},\Sigmabf)$. In the second estimator, the friction vector is modeled as a realization from a GP with mean function $\hat{\mubf}$ and covariance function $\Sigmabf$,
\begin{equation}\label{eq:forceGP}
	\mubf^y(\alphabf(\xest,\uest)) \sim \GP(\hat{\mubf}(\alphabf(\xest,\uest)),\Sigmabf(\alphabf(\xest,\uest))).
\end{equation} 
The difference is that \eqref{eq:stiffness} is a linear model, that is, it well represents the true tire-friction curve during normal driving conditions but does not capture aggressive maneuvering. On the other hand, \eqref{eq:forceGP} targets the full friction curve and therefore has more potential in providing high performance also for aggressive maneuvering, but is more complex and therefore needs more excitation of the driving data. 
\begin{remark}
	\label{rem:UnderEFstimateRemark}
The estimator targeting the tire stiffness uses a model of the tire friction that is linear in the slip angle according to \eqref{eq:stiffness}. This implies that the tire-stiffness estimator will underestimate the available tire friction as larger slip angles are attained (i.e., as more aggressive maneuvering is performed). Fig.~\ref{fig:pacejkasat} provides an illustration of why this occurs for an asphalt tire model.
When the vehicle operates at nonzero slip angles, the estimated tire-friction model can be thought of as a line between the origin and the true tire friction.
The slope of this line (the estimated stiffness) decreases as the slip angle increases and the tire-friction curve flattens.
	\begin{figure}
	\input{tikzdefs}
	\centering

	\input{tikztrim}
	\begin{minipage}{\columnwidth} 
				\centering
			
			\newcommand{\pgfigwidth}{0.85\textwidth}
			\newcommand{\pgfigheight}{9cm}
			\tikzsetnextfilename{StiffnessIllustration}
			\filemodCmp{\figdir/StiffnessIllustration.tikz}{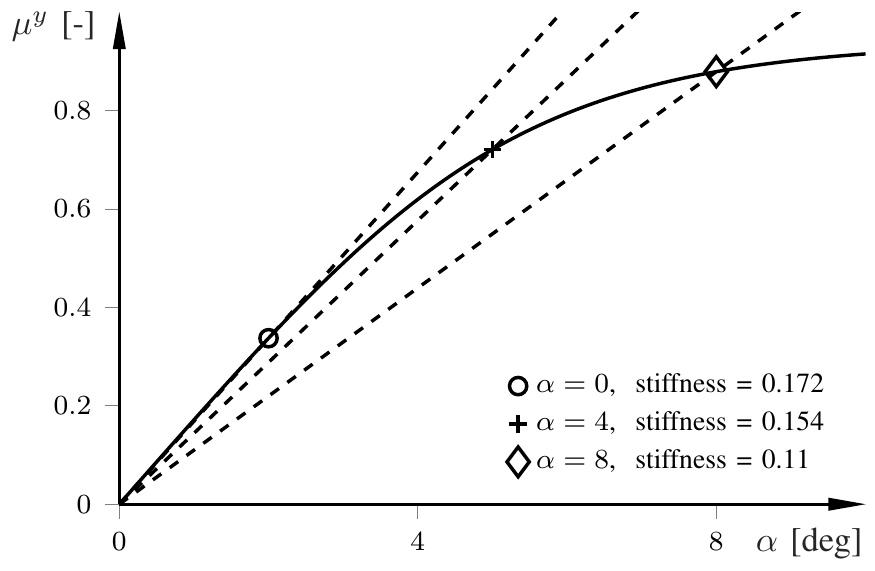}%
			{\tikzcustomremake{StiffnessIllustration}{\figdir\StiffnessIllustration}}
			{}%
			{\input{\figdir/StiffnessIllustration.tikz}}
	\end{minipage}%

		\vspace*{-10pt}
		\caption{Linear approximation of the tire friction at slip angles of 0, 4, 8~deg for an asphalt tire model. The solid line is the true tire-friction curve, while dashed lines are the linear approximations for the respective slip angle. The stiffness (slope of the line going through the origin) decreases as the slip angle increases.}\label{fig:pacejkasat}
		\vspace{-15pt}
	\end{figure}

\end{remark}

Irrespective of using \eqref{eq:stiffness} or \eqref{eq:forceGP}, with a time discretization using the sampling period $T_s$, the vehicle dynamics model used in the estimator can be written as 
\begin{equation}\label{eq:EstModel}
	\xest_{k+1}  = \fvec^e(\xest_k,\uest_k) + \gvec^e(\xest_k,\uest_k)\mubf(\alphabf_k(\xest_k,\uest_k)),
\end{equation}
\red{which is obtained from inserting \eqref{eq:stiffness} or \eqref{eq:forceGP} into \eqref{eq:st1}. In this paper, a simple forward Euler method is used to discretize the dynamics for estimation purposes, but any discretization scheme can be applied.} 
The lateral velocity and yaw rate are estimated as part of the state $\xest$ according to \eqref{eq:st1}. The longitudinal velocity can be determined from the wheel speeds $\{\omega_i\}_{i=f,r}$ and the longitudinal acceleration measurement $a^X$.

Our measurement model is based on a setup commonly available in production cars, namely the acceleration $a^Y$, and yaw-rate measurements $\dot{\psi}$, forming the measurement vector 
$
\yvec  = [a^Y , \dot{\psi}
]^\top
$.
To relate $\yvec_k$ to the estimator state $\xest_k$ at each time step $k$, note that $a^Y$ can be extracted from the right-hand side of \eqref{eq:st1b}, and the last measurement is the yaw rate. We model the measurement noise $\evec_k$ as zero-mean Gaussian distributed noise with covariance $\R$ according to $\evec_k \sim \Ncal(\zerovec,\R)$. This results in the measurement model 
\begin{equation}
	\yvec_k = \hvec(\xest_k,\uest_k)  + \D(\xest_k,\uest_k)\mubf(\alphabf_k(\xest_k,\uest_k)) + \evec_k.\label{eq:measmod}
\end{equation}
The measurement covariance $\R$ is assumed known a priori. This is reasonable, since the measurement noise can oftentimes be determined from prior experiments and data sheets. Hence, the estimation problem consists of estimating the vehicle state $\xest_k$ and friction statistics $\{\hat{\mubf},\Sigmabf\}$, where $\{\hat{\mubf},\Sigmabf\}$ represent the initial slope (the tire stiffness) or the full friction curve, depending on the approach taken.

\subsection{Joint State and Friction-Function Learning}
The vehicle state $\xest_k$ needs to be estimated concurrently with the friction function $\mubf$ at each time step $k$. This is easier to do reliably for moderate slip angles corresponding to normal driving as data gathering in the nonlinear regime is potentially difficult. Furthermore, the sensors are automotive grade and therefore have significant noise. Hence, to properly account for the inherent uncertainty and to be able to integrate properly with SNMPC, we approach the estimation problem in a Bayesian framework. The estimation problem is highly non-Gaussian, which is further amplified when we model the tire-friction model using a GP. We therefore leverage a particle-filter based estimator that estimates the vehicle state $\xest$ and friction function $\mubf$. The particle filter approximates the joint posterior density 
\begin{equation}\label{eq:posterior}
	p(\mubf_k,\xest_{0:k}|\yvec_{0:k})
\end{equation}	
at each time step $k$, that is, the posterior density function of the friction function $\mubf_k$ and state trajectory $\xest_{0:k} = \{\xest_0,\ldots,\xest_k\}$ from time step $0$ to $k$, given the measurement history $\yvec_{0:k} = \{\yvec_0,\ldots,\yvec_k\}$. We use the standard decomposition of \eqref{eq:posterior} into 
\begin{equation}\label{eq:decomposition}
	p(\mubf_k,\xest_{0:k}|\yvec_{0:k}) = p(\mubf_k|\xest_{0:k},\yvec_{0:k}) \, p(\xest_{0:k}|\yvec_{0:k}).
\end{equation}	
The two densities on the right-hand side of \eqref{eq:decomposition} can be estimated recursively. We estimate the state trajectory density $p(\xest_{0:k}|\yvec_{0:k})$ by a set of $N_{\mathrm{PF}}$ weighted state trajectories as 
\begin{equation}
	p(\xest_{0:k}|\yvec_{0:k}) \approx \sum\limits_{i=1}^{N_{\mathrm{PF}}} q_k^i \, \delta_{\xesti_{0:k}}(\xest_{0:k}),\label{eq:PF}
\end{equation}
where $q^i_k$ is the weight of the $i^\text{th}$ state trajectory $\xesti_{0:k}$ and $\delta(\cdot) $ is the Dirac delta. 
Given the state trajectory, we can compute the sufficient statistics necessary to approximate $p(\mu_k|\xesti_{0:k},\yvec_{0:k})$ for each particle. Because we determine the state trajectory from the particle filter, the computations leading up to the estimation of $p(\mubf_k|\xesti_{0:k},\yvec_{0:k})$ are analytic.

To determine the covariance and mean function, we can marginalize out the state trajectory from $p(\mubf_k|\xesti_{0:k},\yvec_{0:k})$ according to 
\begin{equation}\label{eq:postfun}
	p(\mubf_k|\yvec_{0:k}) = \int \! p(\mubf_k|\xest_{0:k},\yvec_{0:k})p(\xest_{0:k}|\yvec_{0:k}) \, \der \xest_{0:k}  \approx \sum\limits_{i=1}^{N_{\mathrm{PF}}} q^i_k \, p(\mubf_k|\xesti_{0:k},\yvec_{0:k}),
\end{equation}
from where the mean function $\hat \mubf_k(\alphabf_k(\xest_k,\uest_k))$ and covariance function $\Sigmabf_k(\alphabf_k(\xest_k,\uest_k))$ needed in the SNMPC problem~\eqref{eq:ocp} can be extracted.
\subsection{Computationally Efficient Particle-Filter Implementation}

To enable real-time implementation of the particle filter using automotive-grade sensors, we rely on the marginalization concept, implying that parts of the estimation problem is analytic. For the case of tire-stiffness estimation, by modeling the unknown friction slope in \eqref{eq:stiffness} as a Gaussian random variable according to $\Delta \C \sim \Ncal(\hat{\C},\Sigmabf)$, we can estimate the state trajectory with a particle filter, and the sufficient statistics of the tire-stiffness parameters are subsequently updated analytically by relying on suitable \emph{conjugate priors} \cite{Berntorp2018h}.

On the other hand, for the GP-based friction estimator, a bottleneck in some of the proposed GP-SSM methods is the computational load. In this paper, we use a computationally efficient reduced-rank GP-SSM framework presented in \cite{Solin2020,Svensson2017}, where the GP is approximated by a basis function expansion using the Laplace operator eigenfunctions
\begin{equation}\label{eq:phi_1}
		\phi^j(x)  = \frac{1}{\sqrt{L}}\sin{\left(\frac{\pi j(x+L)}{2L}\right)},
\end{equation} 
defined on the interval $[-L,L]$, such that 
\begin{equation}\label{eq:BSE}
	\mu^y_i \approx \sum_j \gamma^j_{i} \, \phi^j(\alpha_i),
\end{equation} where the weights $\gamma^j_{i}$ are Gaussian random variables with unknown mean and covariance, whose prior depends on the spectral density that is a function of the eigenvalues in \eqref{eq:phi_1}. For convenience, we express the prior on the coefficients $\gamma_{i}^j$ at time step $k=0$ as a zero-mean matrix-normal ($\MN$) distribution over $\A$, 
\begin{equation}\label{eq:MN}
	\A\sim \MN(\zerovec,\Q,\V),\end{equation} with right covariance $\Q$ and left covariance $\V$. We can write the basis-function expansion in matrix form as 
\begin{equation}\label{eq:GP-SSM}
	\mubf = \underbrace{\begin{bmatrix}
			\gamma_f^{1} & \cdots & \gamma^{m}_f & \zerovec & \cdots & \zerovec\\\zerovec & \cdots & \zerovec & \gamma^{1}_{r} & \cdots & \gamma_{r}^{m}
	\end{bmatrix}}_{\A=\begin{bmatrix} \A_f &  \zerovec\\ \zerovec & \A_r\end{bmatrix}}\underbrace{\begin{bmatrix}
			\phi^{1}(\alpha_{f})\\ \vdots \\ \phi^{m}(\alpha_{f}) \\ 		\phi^{1}(\alpha_{r})\\ \vdots \\ \phi^{m}(\alpha_{r})
	\end{bmatrix}}_{\varphibf(\alphabf)},
\end{equation} where $\gamma^j_i$ are the weights to be learned and $m$ is the total number of basis functions. This formulation leads to a computationally efficient marginalized particle-filter implementation where, similar to the linear case, we can update the sufficient statistics analytically. 
\begin{remark}
	According to well-established tire models (e.g.,~\cite{Pacejka2006}, see Fig.~\ref{fig:pacejka}), it is known that the tire-friction function is antisymmetric. At the same time, the basis-function expansion \eqref{eq:BSE} is a weighted sum of sinusoids for each dimension. Hence, we can trivially enforce antisymmetry by restricting the sum in~\eqref{eq:BSE} to even values for $j$ (i.e., $j=2,4,\ldots,m$). This substantially reduces the number of parameters to estimate and at the same time ensures that the estimated friction function \eqref{eq:forcedef} passes through the origin. 
\end{remark}
\begin{remark}
	Because of the approximation of the GP as a basis-function expansion, the infinite-dimensional function learning problem is reduced to a finite-dimensional parameter learning problem. Hence, in the end, both methods, while algorithmically slightly different, conceptually both estimate parameters.
\end{remark}

\section{STOCHASTIC NONLINEAR MODEL PREDICTIVE CONTROL FOR REAL-TIME VEHICLE CONTROL}
\label{sec:SNMPC}
Based on~\eqref{eq:st1}--\eqref{eq:alphalambda} and using a \red{fourth-order Runge-Kutta method} with sampling period $T_s$, the complete nonlinear vehicle model used for control can be written in the form 
\begin{equation}
	\xvec_{k+1} = \fvec(\xvec_k, \uvec_k, \mubf(\xvec_k,\uvec_k)), \label{eq:sysDyn}
\end{equation}
where $\xvec_k \in \Real^{n_\mathrm{x}}$ denotes the state, $\uvec_k \in \Real^{n_\mathrm{u}}$ the control inputs, and $\mubf(\xvec_k,\uvec_k)$ is the tire-friction function, which is the disturbance (the process noise) that we estimate at each time step $k$ using the methods described in Sec.~\ref{sec:BL}. The state and control vector in the SNMPC formulation are
\begin{equation}
	\begin{aligned}
		\xvec &:= \begin{bmatrix} p^{\bf X},\; p^{\bf Y},\; \psi,\; v^X,\; v^Y,\; \dot{\psi},\; \delta_f\end{bmatrix}^\top, \\
		\uvec &:= \begin{bmatrix} \dot{\delta}_f,\; \omega_f, \;\omega_r\end{bmatrix}^\top,
	\end{aligned}
	\label{eq:state_ctrl}
\end{equation}
such that $n_{\mathrm{x}} = 7$ and $n_{\mathrm{u}} = 3$.

At each sampling time, based on the given state estimate $\hat{\xvec}_t$ and covariance $\PP_t$ at the current time step $t$, the SNMPC solves the following tracking-type OCP 
\begin{subequations}
	\label{eq:ocp}
	\begin{alignat}{2}
		\min_{\xvec_k, \uvec_k,\PP_k}\; & 
		\sum_{k=0}^{N_\mathrm{MPC}-1}\, l_k(\xvec_k, {\uvec}_k) + l_{N_\mathrm{MPC}}(\xvec_{N_\mathrm{MPC}})\label{eq:ocp1}\\
		\text{s.t.} \quad &
		\xvec_0 = \hat{\xvec}_t, \quad \PP_0 = \PP_t, \label{eq:ocp5}\\
		& \forall k \in \{0, \dots, N_\mathrm{MPC}-1\}:\nonumber\\
		& \xvec_{k+1} = \fvec(\xvec_k, {\uvec}_k,\hat{\mubf}_t(\xvec_k,{\uvec}_k)), \label{eq:ocp4}\\
		& \PP_{k+1} = \F_k \PP_k \F_k^\top + \G_k \Sigmabf_t(\xvec_k,{\uvec}_k) \G_k^\top,  \label{eq:ocp3}\\
		&Pr\left( \mathbf{c}(\xvec_k, {\uvec}_k) \le 0 \right)\geq 1-\epsilon, \label{eq:ocp2}
	\end{alignat}
\end{subequations}
where the control action is in the feedforward-feedback form ${\uvec}_k = \uvec_{\mathrm{ref},k} + K(\xvec_k-\xvec_{\mathrm{ref},k}) + \Delta \uvec_k$ due to a prestabilizing controller, \red{where $\uvec_{\mathrm{ref},k}$ and $\xvec_{\mathrm{ref},k}$ are provided by a reference generator, for example, a motion planner~\cite{Paden2016,Berntorp2017c}. The matrices} $\F_k(\cdot)$, $\G_k(\cdot)$ are the Jacobian matrices of \eqref{eq:sysDyn} linearized with respect to $\xvec$ and $\mubf$, respectively, which are functions of the state $\xvec_k$, control ${\uvec}_k$ and mean friction estimate $\hat{\mubf}_t$. The state covariance propagation in~\eqref{eq:ocp3} corresponds to the extended Kalman filter~(EKF) equations to define the state covariance matrix $\PP_k$ for $k=1, \ldots, N_\mathrm{MPC}$.

In linearization-based SNMPC, the model disturbance is Gaussian assumed and state independent, for example, see~\cite{Feng2020}. When we model the friction as a Gaussian random variable, as we do for the estimator targeting the initial slope of the tire-friction curve, the estimator and SNMPC can be straightforwardly connected. However, when the tire-friction function is a GP and therefore state dependent, we need to adapt the covariance propagation \eqref{eq:ocp3} to account for this state dependence by integrating the estimated tire-friction function with the SNMPC (see Sec.~\ref{sec:RCS}). This amounts to modifying the calculation of the Jacobian matrices $\F_k(\cdot)$ and $\G_k(\cdot)$.

\subsection{Objective Function and Inequality Constraints}
In our tracking-type OCP formulation, we consider the stage cost and terminal cost in~\eqref{eq:ocp1} to be the least-squares functions
\begin{equation}
	\begin{aligned}
		l_k(\cdot) &= \frac{1}{2}\|\xvec_k - \xvec_{{\rm ref}, k}\|^2_{Q} + \frac{1}{2}\|{\uvec}_k - \uvec_{{\rm ref}, k}\|^2_{R} + r_s s_k, \\
		l_{N_\mathrm{MPC}}(\cdot) &= \frac{1}{2}\|\xvec_{N_\mathrm{MPC}} - \xvec_{{\rm ref}, N_\mathrm{MPC}}\|^2_{Q_{N_\mathrm{MPC}}}, 
	\end{aligned}
	\label{eq:LScost}
\end{equation}
which includes a term for both state and control reference tracking, and an L1 term for penalizing the slack variable $s_k \ge 0$. The constraints $\mathbf{c}(\xvec_k, \uvec_k) \le 0$ in the OCP~\eqref{eq:ocp} consist of geometric and physical limitations on the system. In practice, it is important to reformulate these requirements as soft constraints, based on the slack variable $s_k$. Otherwise, the problem may become infeasible due to unknown disturbances and modeling errors, and the controller will cease to operate. The constraints in~\eqref{eq:ocp2} therefore include soft bounds on 
\begin{subequations}\label{eq:constraints}
	\begin{align}
		p^{\bf Y}_\text{min}-s_k &\leq p^{\bf Y}_k\leq p^{\bf Y}_\text{max} + s_k, \label{eq:lateral}\\
		-\delta_{f,\text{max}}-s_k &\leq \delta_{f,k}\leq \delta_{f,\text{max}}+s_k, \label{eq:delta}\\
		-\dot \delta_{f,\text{max}}-s_k &\leq \dot \delta_{f,k}\leq \dot \delta_{f,\text{max}}+s_k, \label{eq:deltadot}\\
		\omega_\text{min}-s_k &\leq \omega_{i,k}\leq \omega_\text{max}+s_k,\ \quad i \in \{f,r\}, \label{eq:wheelSpeed} \\
		-\alpha_\text{max}-s_k &\leq \alpha_{i,k}\leq \alpha_\text{max}+s_k,\ \quad i \in \{f,r\}, \label{eq:slipAngle}
	\end{align} 
\end{subequations}
where the slack variable is restricted to be \red{nonnegative} $s_k \ge 0$.
Eq.~\eqref{eq:lateral} bounds the lateral position and is used to ensure that the vehicle stays on the road.
Eqs.~\eqref{eq:delta}-\eqref{eq:slipAngle} bound the wheel angle, inputs, and slip angles, \red{where the slip angles are defined in~\eqref{eq:alphalambda}.
The input constraints in~\eqref{eq:deltadot}-\eqref{eq:wheelSpeed} are imposed as soft constraints, due to the state feedback formulation for the control action ${\uvec}_k = \uvec_{\mathrm{ref},k} + K(\xvec_k-\xvec_{\mathrm{ref},k}) + \Delta \uvec_k$ in~\eqref{eq:ocp}. An affine scaling of the slack variable in the soft constraints~\eqref{eq:constraints} can be used to account for the different dimensions of each of the variables. Using one slack variable $s_k \ge 0$ in each of the soft constraints~\eqref{eq:constraints} corresponds to penalizing the maximum violation of the constraints at each time step. 
The weight value $r_s > 0$ in~\eqref{eq:LScost} for the L1 slack penalty needs to be chosen sufficiently large~\cite{kerrigan2000soft}, such that all slack variables are zero, i.e., $s_k=0$ if a feasible solution exists.
Alternatively, a separate slack variable and separate weight value could be used for each of the constraints. However, this would increase the total number of optimization variables and therefore increase the computational complexity of solving the resulting OCP in general.}

\subsubsection{Stability Constraints for the Tire-Stiffness SNMPC}
As the tire-stiffness estimator targets the linear region of the tire-friction curve, for aggressive maneuvering we need to ensure that the SNMPC does not operate excessively outside the validity region of the estimator. To this end, when the SNMPC is connected to the tire-stiffness estimator, we include the additional constraints
\begin{equation}
| \dot{\psi}_kv_k^x| \leq 0.85\mu_c g, \quad \left|\frac{v_k^y}{v_k^x}\right| \leq \tan^{-1} (0.02 \mu_c g), \label{eq:stab}
\end{equation}
where $\mu_c$ is the friction coefficient. The constraints in~\eqref{eq:stab} prevent the vehicle from entering regions of high lateral acceleration and side slip, and can be found in~\cite[Chapter~8]{Rajamani2006}.
We refer to~\eqref{eq:stab} as \emph{stability constraints}. The stability constraints depend on the (peak) friction coefficient $\mu_c$. Experimental studies suggest that using a monotonic relationship is sufficient to differentiate between asphalt and snow~\cite{Gustafsson1997,Ahn2012}.
In this work, we use a linear relationship to approximate the friction coefficient as a function of the tire-stiffness estimate, 
\begin{align}\label{eq:muapprox}
\mu_c \approx \min\left ( \frac{a(C_{f,n}^y +\Delta C_{f}^y+ C_{r,n}^y+\Delta C_{r}^y)}{2},1\right),
\end{align}
where $a$ is a constant that in this work is adjusted from Pacejka models for asphalt and snow. The central idea of \eqref{eq:muapprox} is that the bounds on the acceleration and sideslip should tighten as the friction coefficient, and consequently the cornering stiffness, decreases. 
For surfaces such as wet asphalt, which may have a high cornering stiffness but lower road friction, \eqref{eq:muapprox} is conservative because the stiffness estimator underestimates the true stiffness as the tires saturate (as in Fig.~\ref{fig:pacejkasat}, see Remark~\ref{rem:UnderEFstimateRemark}).
\subsection{Probabilistic Chance Constraints}
To enforce the probabilistic chance constraints in~\eqref{eq:ocp2}, we reformulate them as deterministic constraints as in~\cite{Telen2015}, where the $j^{\text{th}}$ constraint is written as
\begin{align}
	c_j(\xvec_k, \uvec_k) + \nu \sqrt{\frac{\partial c_j}{\partial \xvec_k}\PP_k\frac{\partial c_j}{\partial \xvec_k}^T} \leq 0,
\end{align}
where $\nu$ is referred to as the back-off coefficient and depends on the desired probability threshold $\epsilon$ and assumptions about the resulting state distribution.
The backoff coefficient for Cantelli's inequality, $\nu = \sqrt{\frac{1-\epsilon}{\epsilon}}$ 
, holds regardless of the underlying distribution but is conservative.
In this work, we approximately assume normal-distributed state trajectories and set
\begin{align}
	\nu = \sqrt{2}\text{erf}^{-1}(1-2\epsilon),
\end{align}
where $\text{erf}^{-1}(\cdot)$ is the inverse error function.

\subsection{Online SNMPC and Software Implementation} \label{sec:software}

The resulting SNMPC implementation, proposed recently in \cite{Feng2020}, uses the QP solver \texttt{PRESAS} \cite{Quirynen2020}, which applies block-structured factorization techniques with low-rank updates to preconditioning of an iterative solver within a primal active-set algorithm with tailored initialization methods. This results in a simple, efficient and reliable QP solver suitable for embedded control hardware. The algorithm uses a tailored Jacobian approximation along with an adjoint-based SQP method that allows for the numerical elimination of the covariance matrices from the SQP subproblem, which reduces the computation time when compared to standard SQP methods applied to SNMPC~\cite{Feng2020}. The nonlinear function and derivative evaluations, for the preparation of each SQP subproblem, are performed using algorithmic differentiation~(AD) and C~code generation in \software{CasADi}~\cite{Andersson2018}. In addition, a standard line search method is used~\cite{Nocedal2006} to improve the closed-loop convergence of the SQP-based SNMPC controller.

We solve the nonlinear OCP \eqref{eq:ocp} only approximately at each control time step by a tailored implementation of the real-time iteration (RTI) scheme \cite{Diehl2005} using the adjoint-based SQP method~\cite{Feng2020}.
RTI performs one SQP iteration per time step of control, and uses a continuation-based warm starting of the state and control trajectories from one time step to the next~\cite{Gros2016}. RTI may not approximate well the OCP if the problem is linearized far from a local minimum. However, as we envision an autonomous-driving system \cite{Berntorp2018d}, the MPC tracks a trajectory generated by a motion planner to be a suitable reference for linearization, e.g., kinematically feasible and constraint aware. Thus, RTI seems a suitable approach for this application.

\begin{remark}
	Due to time delays resulting from vehicle network communication and computation delays, it is beneficial to introduce time-delay compensation. Denote the time delay as $T_d$. Time-delay compensation can be done by defining the estimates at time $t$ entering the OCP \eqref{eq:ocp} to equal the predicted estimates at time $t+T_d$, computed from the estimates at time $t$ and the input signals up until time $t$ \cite{Berntorp2019f,Berntorp2018d}. 
\end{remark}

\section{Adaptation of SNMPC to the Learned Tire-Friction Function}
\label{sec:RCS}
In determining the mean $\hat{\mubf}_t$ and covariance $\Sigmabf_t$ of the tire-friction in~\eqref{eq:ocp}, we use that each particle retains its own estimate $\mubf^i_k$ together with the weight $q^i_k$. Then, we can estimate the mean and covariance by choosing the $i$th particle that fulfills ${i = \arg \max _{i\in[1,N] } q^i_k}$, or we can choose to estimate the mean and covariance as the weighted mean among the particles. 

When integrating the tire-stiffness estimator with the SNMPC, the connection is already there because the tire-stiffness estimator directly outputs the mean and covariance of the tire-stiffness estimate, which is already in the form that the SNMPC admits. However, for the GP case we need to make some adjustments. 
Using the single-track model and the basis-function expansion of the friction, $\mubf \approx \A \varphibf(\alphabf(\xvec))$, we can write the vehicle model~\eqref{eq:sysDyn} used by the SNMPC in the same format as~\eqref{eq:EstModel},
\begin{equation}\label{eq:MPCModel}
	\xvec_{k+1}  = \fvec(\xvec_k,\uvec_k) + \gvec(\xvec_k,\uvec_k)\hat{\mubf}_k(\alphabf(\xvec_k)).
\end{equation}
Note that $\delta_f$ is part of the state in the MPC formulation~\eqref{eq:state_ctrl} and the slip angle $\alphabf(\cdot)$ is therefore a function of only $\xvec$.

In linearization-based SNMPC~\cite{Feng2020}, the parametric uncertainty is typically modeled according to a Gaussian random variable, $\mubf\sim \Ncal(\hat{\mubf},\Sigmabf)$, where the uncertainty $\mubf$ is state independent. This leads to Jacobian matrices $\F_k = \frac{\partial \fvec}{\partial \xvec}(\xvec_k, {\uvec}_k,\hat{\mubf}_k)$ and $\G_k = \frac{\partial \fvec}{\partial \mubf}(\xvec_k, {\uvec}_k,\hat{\mubf}_k)$ in~\eqref{eq:ocp3}. 
However, in our case, the friction uncertainty is a function that is modeled according to a GP at each time step $k$, ${\mubf(\alphabf(\xvec)) \sim \GP\left(\hat{\mubf}_k(\alphabf(\xvec)),\Sigmabf(\alphabf(\xvec))\right)}$. Hence, we want to determine the equivalent to \eqref{eq:ocp3} for the state-dependent uncertainty $\mubf(\alphabf(\xvec))$, by finding expressions for the involved Jacobians $\F_k$, $\G_k$. 
Starting from~\eqref{eq:MPCModel}, by using the chain rule and $\mubf \approx \A \varphibf(\alphabf(\xvec))$,
\begin{align}
	\F_k & = \frac{\partial \fvec}{\partial \xvec}(\xvec_k, {\uvec}_k) + \frac{\partial \gvec }{\partial \xvec}(\xvec_k, {\uvec}_k)\hat{\mubf}_t(\alphabf(\xvec_k)) \nonumber\\ &\quad + \gvec(\xvec_k, {\uvec}_k) \frac{\partial \hat{\mubf}_t(\alphabf(\xvec_k)) }{\partial \xvec}\nonumber\\
	& = \frac{\partial \fvec}{\partial \xvec}(\xvec_k, {\uvec}_k) + \frac{\partial \gvec }{\partial \xvec}(\xvec_k, {\uvec}_k)\hat{\A}_t\varphibf(\alphabf(\xvec_k)) \nonumber\\ &\quad + \gvec(\xvec_k, {\uvec}_k) \hat{\A}_t \frac{\partial \varphibf }{\partial \alphabf}(\alphabf(\xvec_k)) \frac{\partial \alphabf}{\partial \xvec}(\xvec_k).\label{eq:Jacobian}
\end{align}
Hence, for ease of notation, if we momentarily define $\gvec_k := \gvec(\xvec_k,{\uvec}_k)$, $\varphibf_k := \varphibf(\alphabf(\xvec_k))$, we can approximate the covariance propagation as 
\begin{align}
	\Exp[\xvec_{k+1}\xvec_{k+1}^\top] & \approx \Exp[(\F_k \xvec_k + \gvec_k\hat{\A}_t\varphibf_k)(\F_k \xvec_k + \gvec_k\hat{\A}_t\varphibf_k)^\top]\nonumber\\
	& =\F_k\PP_k \F_k^\top + \Exp (\gvec_k\hat{\A}_t\varphibf _k\varphibf_k^\top \hat{\A}_t^\top \gvec_k^\top)\nonumber\\ & =
	\F_k \PP_k \F_k^\top + \underbrace{\gvec_k}_{\G_k}\underbrace{\cov (\hat{\A}_t\varphibf_k)}_{\Sigmabf_k} \gvec_k^\top. \label{eq:covProp}
\end{align}
Eq.~\eqref{eq:covProp} leads to the SNMPC formulation, where the integration of the tire-friction estimates is done by using the estimated mean in the state prediction~\eqref{eq:ocp4} and the estimated covariance function through the Jacobians in the covariance propagation~\eqref{eq:ocp3}. Algorithm~\ref{alg:1} summarizes our proposed friction-adaptive SNMPC strategy.
\begin{algorithm}
	\caption{Proposed Real-Time SNMPC with Friction Adaptation}
	\label{alg:1}
	\begin{algorithmic}[1]
		\For{each time step $k$}
		\State \parbox[t]{\dimexpr .94\linewidth-\algorithmicindent}{Estimate current state vector $\hat{\xvec}_k$, 
			tire-friction estimate $\hat{\mubf}_k(\alphabf(\xvec^e_k,\uvec^e_k))$ and covariance function $\Sigmabf_k(\alphabf(\xvec^e_k,\uvec^e_k))$, using either of \cite{Berntorp2018h} or \cite{Berntorp2021d}.} \vspace{2mm}
		\State \parbox[t]{\dimexpr .94\linewidth-\algorithmicindent} {Perform SQP iteration(s) to approximately solve SNMPC problem~\eqref{eq:ocp}, using state estimate in~\eqref{eq:ocp5}, mean friction function in~\eqref{eq:ocp4}, and covariance propagation in~\eqref{eq:ocp3} based on \eqref{eq:Jacobian}-\eqref{eq:covProp} for tire-friction estimation.}
		\EndFor
	\end{algorithmic}
\end{algorithm}

\section{Friction-Adaptive Simulation Results}
\label{sec:res}
In this section, we validate the proposed friction-adaptive SNMPC in simulation and in a real-time computing environment. In the simulation studies, we consider a sequence of nine double lane-change maneuvers similar to the standardized IS0 3888-2~\cite{Iso2002} double lane-change maneuver commonly used in vehicle stability tests, with the middle three on snow and the rest on dry asphalt. To investigate the learning behavior of the controllers, the surface change occurs during a straight portion, where the friction curve is unobservable. The reference is generated with Bezier polynomials and the position, heading, and longitudinal velocity are given to the controllers to track. For simplicity, we set the velocity reference to be constant and evaluate the proposed control strategy for two different reference velocities. 
\red{First, we evaluate different combinations of controllers and estimators in Matlab simulations using the same vehicle model for both generating synthetic data and the controller, but with different friction models. Second, we evaluate the controllers in simulation using CarSim~\cite{carsim}. Third, we} verify the real-time feasibility by executing the proposed method in a dSPACE MicroAutoBox-II equipped with an IBM PPC 900MHz processor and $16$ MB main memory, which closely resembles the capabilities of current and near-future embedded microcontrollers
for automotive applications.

The vehicle parameters are from a mid-size SUV, and the tire parameters for the different surfaces are taken from \cite{Berntorp2014a}. The simulation model is the nonlinear single-track model \eqref{eq:st1} with a Pacejka tire model \cite{Pacejka2006} with the friction ellipse for modeling combined slip,
\begin{equation}
			\begin{aligned}
			F_i^x &= \mu_i^xF_i^z\sin(D_i^x\arctan(B_i^x(1-E_i^x)\lambda_i+E_i^x\arctan(B_i^x\lambda_i))), \\
			F_i^y &= \eta_i\mu_i^yF_i^z\sin(D_i^y\arctan(B_i^y(1-E_i^y)\alpha_i+E_i^y\arctan(B_i^y\alpha_i))), \\
			\eta_i&=\sqrt{1-\left(\frac{F_i^x}{\mu_i^xF_i^z}\right)^2},
		\end{aligned} \label{eq:Pacejka}
\end{equation} 
where $\mu_i^j$, $B_{i}^j$, $D_{i}^j$ and $E_{i}^j$, for $i \in \{f,r\}, j \in \{x,y\}$, are the friction coefficients and stiffness, shape, and curvature factors, with values from \cite{Berntorp2014a}. The noise levels in the measurement model \eqref{eq:measmod} are taken from those of a low-cost inertial measurement unit typical for automotive applications. The estimators use $N_\mathrm{PF} = 100$ particles and the tire-friction estimator uses $m=10$ basis functions, five in each direction by exploiting antisymmetry. The initial estimates and the different tuning parameters
in the estimator are fairly generic, and the same as in \cite{Berntorp2020a}. 

We evaluate the following seven controllers:
\begin{itemize}
	\item \textsc{Friction-SNMPC}: the proposed adaptive SNMPC method (Algorithm~\ref{alg:1}) with tire-friction estimator from \cite{Berntorp2021d}.
	\item \textsc{Stiffness-SNMPC}: the proposed adaptive SNMPC method (Algorithm~\ref{alg:1}) with tire-stiffness estimator from~\cite{Berntorp2018h}.
	\item \textsc{Friction-NMPC}: Nominal adaptive NMPC with the tire-friction estimator in~\cite{Berntorp2021d}, i.e., it does not use the covariance propagation and chance constraints.
	\item \textsc{Stiffness-NMPC}: Nominal adaptive NMPC with the tire-stiffness estimator in~\cite{Berntorp2018h}, i.e., it does not use the covariance propagation and chance constraints.
	\item \textsc{Snow-NMPC}: Nominal NMPC that uses fixed friction coefficient values in the Pacejka tire model~\eqref{eq:Pacejka} corresponding to a snow surface.
	\item \textsc{Asphalt-NMPC}: Nominal NMPC that uses fixed friction coefficient values in the Pacejka tire model~\eqref{eq:Pacejka} corresponding to an asphalt surface.
	\item \textsc{Oracle}: Nominal NMPC with the exact nonlinear tire-friction Pacejka model. This controller acts as ground truth and cannot be implemented in practice.
\end{itemize} We include the \textsc{Oracle} controller to provide a lower bound on cost and constraint violations for the simulations.
Its performance cannot be achieved in practice because it is given the exact tire force curve used by the simulation model; in reality, there will be model mismatch due to inaccuracies in both the tire force and single-track vehicle models. All of the controllers perform 1 SQP iteration per planning instance. The estimators are executed at 100Hz and the different MPCs at 20Hz, with a prediction horizon of 2s. The constraint satisfaction probability for the SNMPCs (\textsc{Friction-SNMPC} and \textsc{Stiffness-SNMPC}) are set to 95\%, i.e., $\epsilon = 0.05$. The least-squares cost~\eqref{eq:LScost} prioritizes the lateral position and wheel speed inputs. The metrics we use to evaluate the controllers are cost and score, and are computed as 
\begin{align}\label{eq:costmetric}
	\text{Cost}&=\sum_k l(\xvec_k,\uvec_k), \\
	\text{Score}&=\sum_k((y_k-y_{\max})_++(y_{\min}-y_k)_+)\, T_s,
\end{align}
summed over the simulation time, where $(\cdot)_+=\text{max}(\cdot,0)$ and $l(\cdot)$ corresponds to the stage cost in the MPC objective function.
\begin{remark}
	In all of the simulations for the considered maneuver, \textsc{Asphalt-NMPC} diverged after the transition to snow and is therefore omitted from the results section.
\end{remark}

\subsection{Closed-loop Estimation Results}
\begin{figure}
	\input{tikzdefs}

	\input{tikztrim}
	\begin{minipage}{\columnwidth} 
				\centering
			
			\newcommand{\pgfigwidth}{0.85\textwidth}
			\newcommand{\pgfigheight}{9cm}
			\tikzsetnextfilename{3DBL3}
			\filemodCmp{\figdir/3DBL3.tikz}{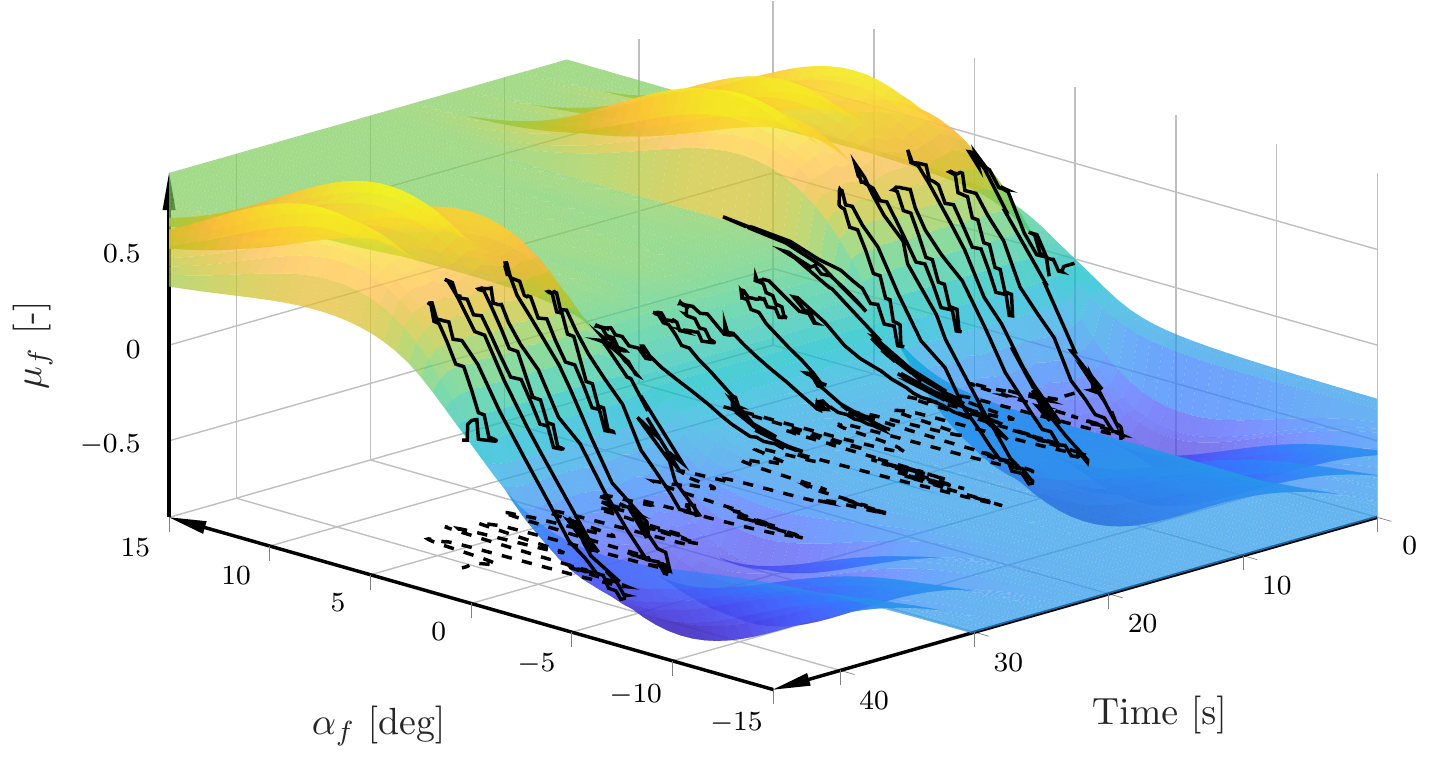}%
			{\tikzcustomremake{3DBL3}{\figdir\3DBL3}}
			{}%
			{\input{\figdir/3DBL3.tikz}}
	\end{minipage}%

	\vspace*{-15pt}
	\caption{Estimated tire-friction function of the front wheel for one realization in closed-loop simulation with multiple double lane-change maneuvers and surface changes for $v_\mathrm{ref}=22$m/s. The trace of $\alpha_f$ is indicated in black and its projection to the $t$-$\alpha_f$ plane is in black dashed.}\label{fig:BLEst1}
	\vspace*{-10pt}
\end{figure}

Fig.~\ref{fig:BLEst1} shows the front tire-friction estimates using the proposed GP-based estimator in Algorithm~\ref{alg:1} in closed loop for one realization of the considered maneuver, with the resulting $\alpha_f$. The estimates are initialized for the parameters on snow. As more measurements are gathered, the estimates converge. The changes from asphalt to snow and vice versa are clearly indicated by the estimator reacting to the abrupt changes. At the switch from asphalt to snow ($t\approx 13$s), the short transient behavior of the estimator can be seen from the slightly larger slip angles around just after the first lane change is initiated on the new surface. However, the increased slip angles result in the estimator learning about the new nonlinearity, and the controller quickly adapts. 

\begin{figure}
	\input{tikzdefsdown}

	\input{tikztrim}
	\begin{minipage}{\textwidth} 
		\begin{flushright}
			
			\newcommand{\pgfigwidth}{0.85\textwidth}
			\newcommand{\pgfigheight}{9cm}
			\tikzsetnextfilename{simeBL}
			\filemodCmp{\figdir/simeBL.tikz}{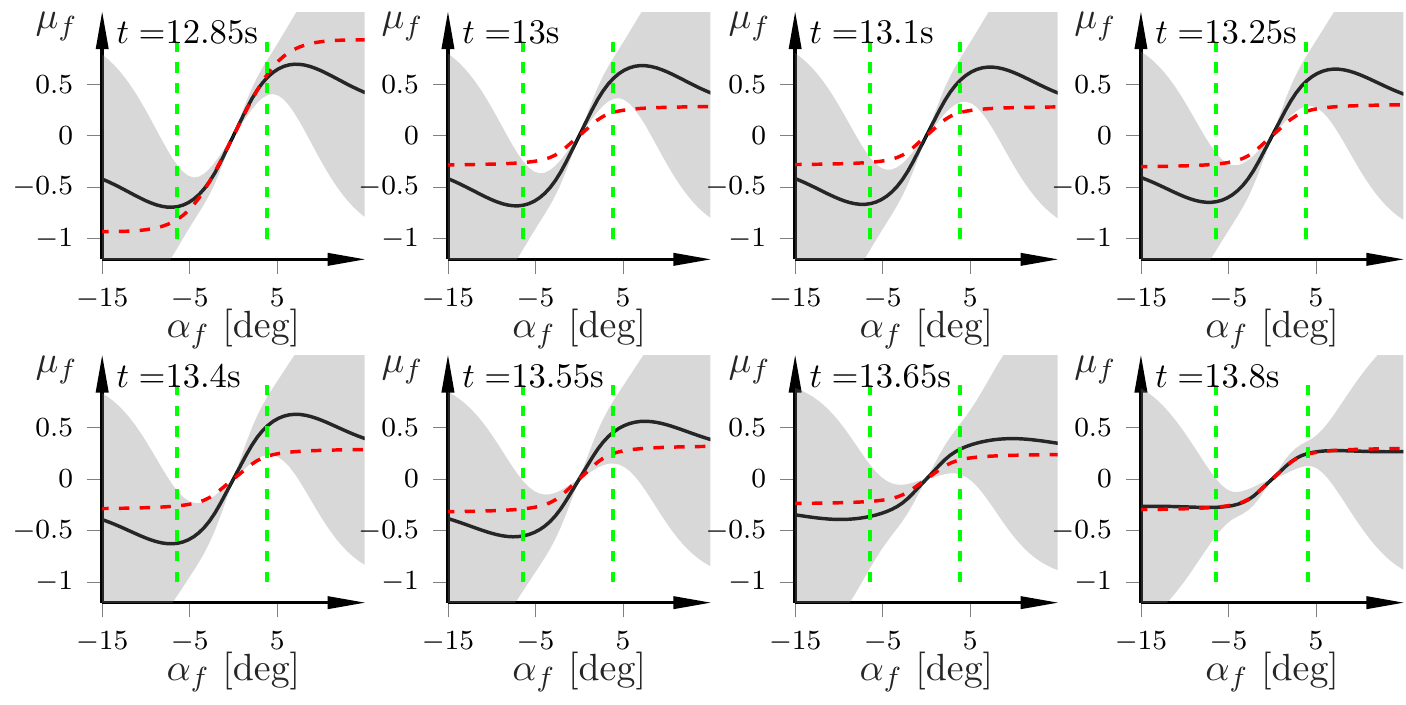}%
			{\tikzcustomremake{simeBL}{\figdir\simeBL}}
			{}%
			{\input{\figdir/simeBL.tikz}}
		\end{flushright}
	\end{minipage}%

	\vspace*{-5pt}
	\caption{The tire-friction estimates (black solid) and estimated $2\sigma$ confidence (gray area) at different time steps for one realization in closed-loop simulation with multiple double lane-change maneuvers and surface changes (see Fig.~\ref{fig:BLEst1} for whole realization). The true tire-friction function is in red dashed, and the estimated range of the $\alpha$ values that have been excited are indicated by the green dashed vertical lines. An abrupt surface change from asphalt to snow occurs right before $t=13$s, and the estimator converges in less than $1$s.}\label{fig:BLEst}
	\vspace*{-10pt}
\end{figure}

To give a more quantitative measure of how the estimator performs, Fig.~\ref{fig:BLEst} shows the results for the same realization based on different snapshots, when an abrupt surface change from dry asphalt to snow is about to occur. For the region of the slip angle corresponding to where data have been acquired, the estimates follow closely to the true friction model, and it is clear that where small amounts of data have been gathered, for example, for slip angles close to or beyond the peak, the uncertainty increases accordingly. The intuition is that such increase of uncertainty can be utilized by the SNMPC when determining the appropriate control commands. The estimator is quick to adapt to the surface change and converges within roughly $1$s. Note, however, that even though it takes about $1$s for the estimator to fully converge, the change in surface is reflected in the uncertainty estimates faster than that.

\begin{figure}
\input{tikzdefs}

	\input{tikztrim}
	\begin{minipage}{\textwidth} 
		\begin{flushright}
			
			\newcommand{\pgfigwidth}{0.85\textwidth}
			\newcommand{\pgfigheight}{6cm}
			\tikzsetnextfilename{stiffness}
			\filemodCmp{\figdir/stiffness.tikz}{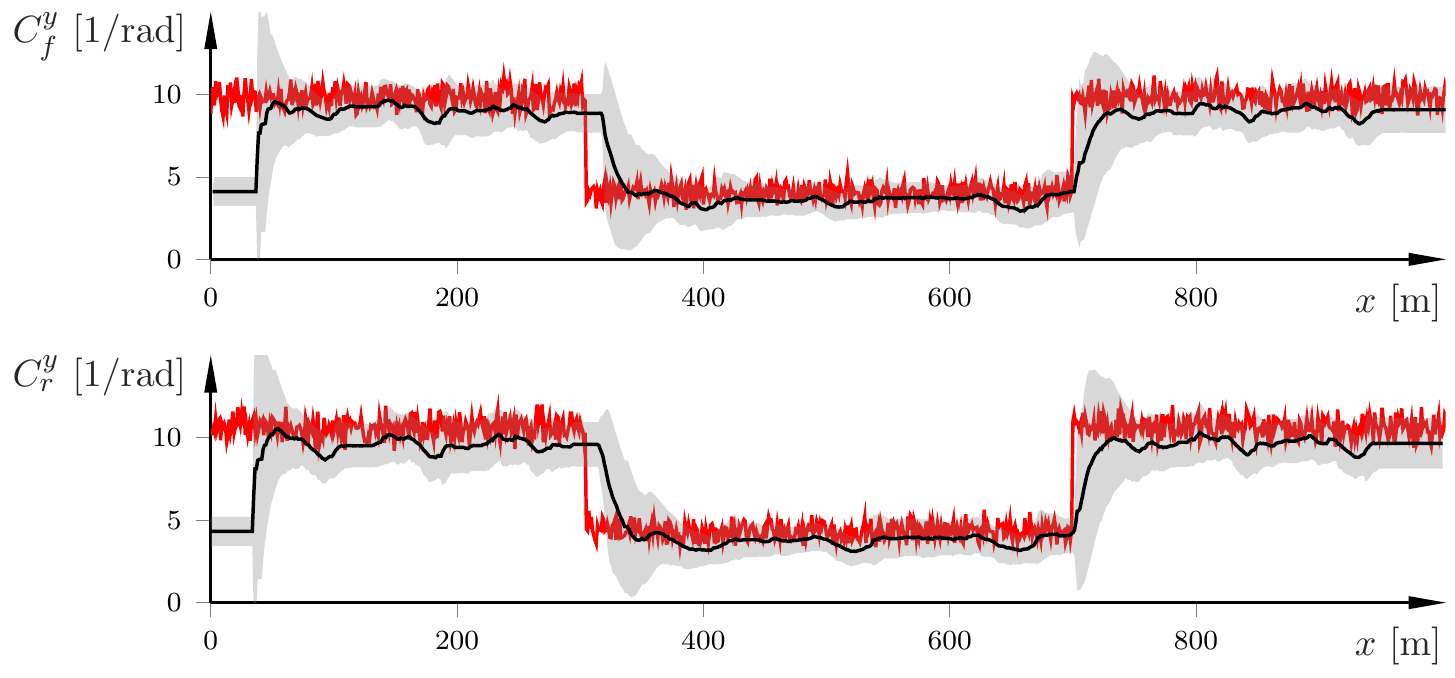}%
			{\tikzcustomremake{stiffness}{\figdir\stiffness}}
			{}%
			{\input{\figdir/stiffness.tikz}}
		\end{flushright}
	\end{minipage}%

	\caption{Estimated tire stiffness for one realization in closed-loop simulation with multiple double lane-change maneuvers and surface changes. The red line is the true stiffness, the slope of the nonlinear tire-friction curve at $\alpha_i=0$. The black solid and gray area are the mean and $2\sigma$~confidence interval from the stiffness estimator, respectively. The stiffness is underestimated when the tire friction saturates, \red{see Remark~\ref{rem:UnderEFstimateRemark}}. The true stiffness corresponds to the initial slopes in Fig.~\ref{fig:BLEst}.}\label{fig:snowstiff}
	\vspace{-5pt}
\end{figure}

Similarly, Fig.~\ref{fig:snowstiff} displays the stiffness estimates for the tire-stiffness estimator in Algorithm~\ref{alg:1}. In this closed-loop simulation, for $v_\mathrm{ref}=22$m/s, the nonlinear region of the tire-friction curve is excited. From the discussion in Remark~\ref{rem:UnderEFstimateRemark}, this leads to an underestimation of the stiffness. However, the true value is still covered in the 95\% confidence intervals. From previous papers, when driving in the linear region, the estimates are highly accurate \cite{Berntorp2018h}.
\subsection{Closed-loop Control Results}
\label{sec:ControlResults}
\begin{figure*}
	\input{tikzdefs}

	\input{tikztrim}
	\begin{minipage}{\textwidth} 
		\begin{flushright}
			
			\newcommand{\pgfigwidth}{0.85\textwidth}
			\newcommand{\pgfigheight}{9cm}
			\tikzsetnextfilename{traj}
			\filemodCmp{\figdir/traj.tikz}{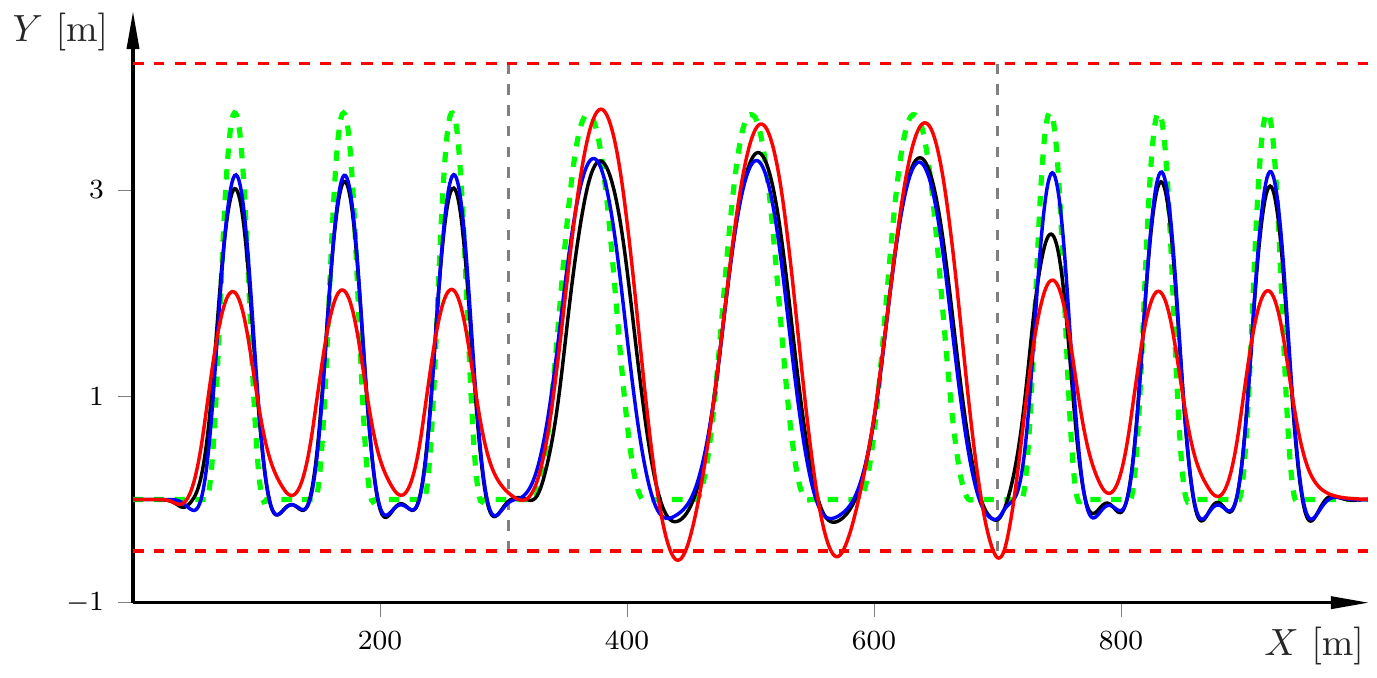}%
			{\tikzcustomremake{traj}{\figdir\traj}}
			{}%
			{\input{\figdir/traj.tikz}}
		\end{flushright}
	\end{minipage}%

	\vspace*{-15pt}
	\caption{Resulting path for the various approaches during the lane-change maneuvers. Red and green dashed lines are the constraints and reference, respectively. The gray vertical dashed lines indicate the surface changes from \red{asphalt-snow-asphalt}. Color coding: \red{\textsc{Friction-SNMPC} (black), \textsc{Oracle} (blue), and \textsc{Snow-NMPC} (red)}.}\label{fig:SNMPCRes}

\end{figure*}

\red{To illustrate the reference path of the maneuver with surface changes, Fig.~\ref{fig:SNMPCRes} shows the trajectories of \textsc{Friction-SNMPC}, \textsc{Snow-NMPC}, and \textsc{Oracle} when setting the velocity reference to $v_\mathrm{ref} = 22$m/s.}
The results demonstrate that the reference trajectory is tracked well by the adaptive \textsc{Friction-SNMPC} controller and the \textsc{Oracle} controller. 
The \textsc{Snow-NMPC} controller performs conservatively in the asphalt sections, and also has difficulty tracking the reference in the snow portion; due to the aggressive nature of the maneuver and inaccuracies between the MPC and simulation model.
\red{After the estimator transients, \textsc{Friction-SNMPC} seems to track the path slightly better. Note that \textsc{Oracle}, which is an ideal controller that cannot be implemented in practice, does not seem to clearly outperform the proposed adaptive controller.}

To numerically quantify the performance and show the merits of the proposed methods, we have executed two sets of 200 Monte-Carlo runs for the considered maneuver, with the velocity reference $v_\mathrm{ref}=19$m/s and $v_\mathrm{ref}=22$m/s, respectively. \red{For each Monte-Carlo simulation, the Pacejka parameters for each road surface are randomly perturbed, with samples drawn from a uniform distribution up to $\pm 10\%$ on asphalt and $\pm20\%$ on snow. In addition, the measurements are subject to zero-mean Gaussian noise with covariance values chosen based on specifications from automotive-grade sensors.} 
Table~\ref{tab:1}~and~\ref{tab:2} show the results.
Clearly, when comparing the friction-adaptive controllers, the stochastic formulations give substantially less constraint violations than their nominal counterparts; for example, the combination of SNMPC with GP-based tire-friction estimation (\textsc{Friction-SNMPC}) leads to constraint satisfaction at all times. Notably, the nominal NMPC with tire-friction estimation (\textsc{Friction-NMPC}) destabilizes for large velocities (Table~\ref{tab:2}), but this does not happen for the nominal counterpart using tire-stiffness estimation. Insight into this is given by inspection of Figs.~\ref{fig:BLEst}~and~\ref{fig:snowstiff}. The tire-stiffness estimator gives an underestimation of the actual stiffness when exciting the nonlinear region of the tire-friction curve, \red{see Remark~\ref{rem:UnderEFstimateRemark}}. Hence, the effect of having slightly biased estimates, is that the controller acts conservatively and therefore implicitly injects robustness into the approach. 
\begin{table} %
	\centering
	\caption{Results for 200 Monte-Carlo runs with $v_\mathrm{ref}=19$m/s.}\label{tab:1}
	\vspace*{-5pt}
	\begin{tabular}{lcccc}
		\toprule %
		Method & Mean Cost & Max Cost & Mean Score & Max Score \\
		\midrule
		\textsc{Friction-SNMPC} & 0.938& 1.008 & 0 & 0 \\
		\textsc{Stiffness-SNMPC} & 1.254& 2.058 & 0.008 & 0.097 \\
		\textsc{Friction-NMPC} &  0.917&0.951 & 0 & 0 \\
		\textsc{Stiffness-NMPC} &  1.740& 4.929 & 0.018 & 0.071 \\
		\textsc{Snow-NMPC} & 3.332& 3.439 &  0.035 & 0.114 \\
		\textsc{Oracle} & 0.710& 0.725 & 0 & 0 \\
		\bottomrule
	\end{tabular}
	\vspace*{-12pt}
\end{table}

\begin{table} %
	\centering
	\caption{Results for 200 Monte-Carlo runs with $v_\mathrm{ref}=22$m/s.}\label{tab:2}
	\vspace*{-5pt}
	\begin{tabular}{lcccc}
		\toprule %
		Method & Mean Cost & Max Cost & Mean Score & Max Score \\
		\midrule
		\textsc{Friction-SNMPC} & 2.187& 2.868 & 0 & 0 \\
		\textsc{Stiffness-SNMPC} & 2.716& 3.098 & 0.002 & 0.055 \\
		\textsc{Friction-NMPC} &  557&72,626 & 22 & 2,973 \\
		\textsc{Stiffness-NMPC} &  2.833& 5.90 & 0.006 & 0.0677 \\
		\textsc{Snow-NMPC} & 4.391& 5.05 & 0.077 & 0.111 \\
		\textsc{Oracle} & 1.682& 1.709 & 0 & 0 \\
		\bottomrule
	\end{tabular}
	\vspace*{-12pt}
\end{table}

In summary, the tire-stiffness estimator gives a good compromise in the trade-off between cost and score, whereas leveraging the potential of full tire-friction curve estimation in an SNMPC formulation gives the best performance. However, care needs to be taken in the design of the MPC used in combination with estimating the full tire-friction curve, as such an estimator demands more from the employed controller.

Finally, in our evaluation, we show the stability constraints \eqref{eq:stab} for \textsc{Stiffness-SNMPC} in the considered realization. As the surface shifts from asphalt to snow, the constraints tighten to maintain performance of the control strategy. These additional constraints turned out to be important for the adequate functioning when estimating a linear tire-friction curve. 
\begin{figure}
	\input{tikzdefs}

	\input{tikztrim}
	\begin{minipage}{\textwidth} 
		\begin{flushright}
			
			\newcommand{\pgfigwidth}{0.85\textwidth}
			\newcommand{\pgfigheight}{6cm}
			\tikzsetnextfilename{StabCons}
			\filemodCmp{\figdir/StabCons.tikz}{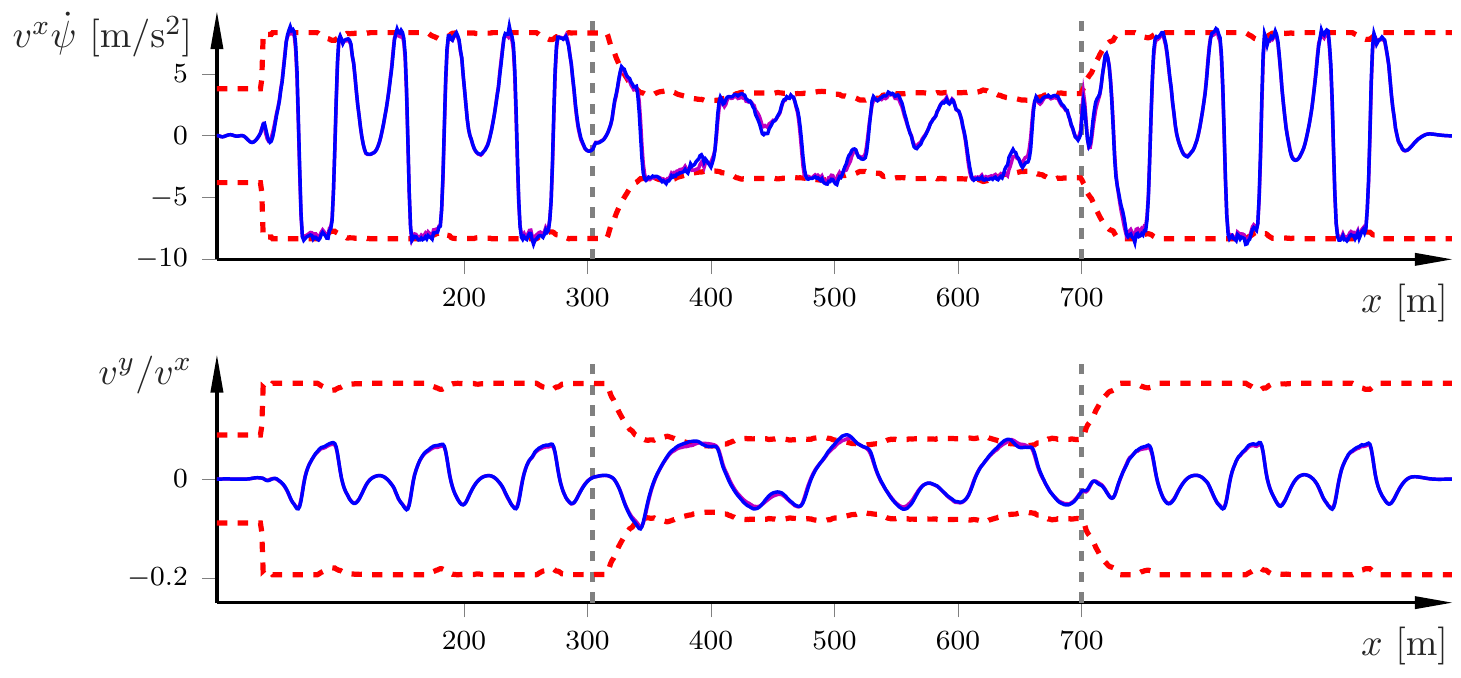}%
			{\tikzcustomremake{StabCons}{\figdir\StabCons}}
			{}%
			{\input{\figdir/StabCons.tikz}}
		\end{flushright}
	\end{minipage}%

	\caption{Stability constraints enforced for the controllers using the stiffness estimator (\textsc{Stiffness-SNMPC} in magenta and \textsc{Stiffness-NMPC} in blue), where the middle portion is on snow. Red dashed lines are the constraint boundaries, where the road friction is calculated with~\eqref{eq:muapprox} using the estimator output. The constraints tighten during the snow portion. Although the differences are minor in this particular realization, \textsc{Stiffness-NMPC} has more constraint violations.}\label{fig:snowcons}
	\vspace{-5pt}
\end{figure}

\red{\subsection{CarSim Validation Results}}
\red{In this section, we further validate the method using the high-fidelity vehicle simulator CarSim. From Tables~\ref{tab:1}~and~\ref{tab:2}, it is clear that \textsc{Friction-SNMPC} is the method that performs best according to the metrics used. Consequently, we choose to evaluate this control system in CarSim and compare with nominal controllers. For reproducibility, we use modules that are standard in CarSim for all components such as suspension, springs, tires, and so on, and the vehicle parameters are from a mid-size SUV. We set the longitudinal velocity reference to be constant, $v^X_\mathrm{ref}=18$m/s.}

\begin{figure}
	\input{tikzdefs}

	\input{tikztrim}
	\begin{minipage}{\textwidth} 
		\begin{flushright}
			
			\newcommand{\pgfigwidth}{0.85\textwidth}
			\newcommand{\pgfigheight}{9cm}
			\tikzsetnextfilename{CarSimTraj}
			\filemodCmp{\figdir/CarSimTraj.tikz}{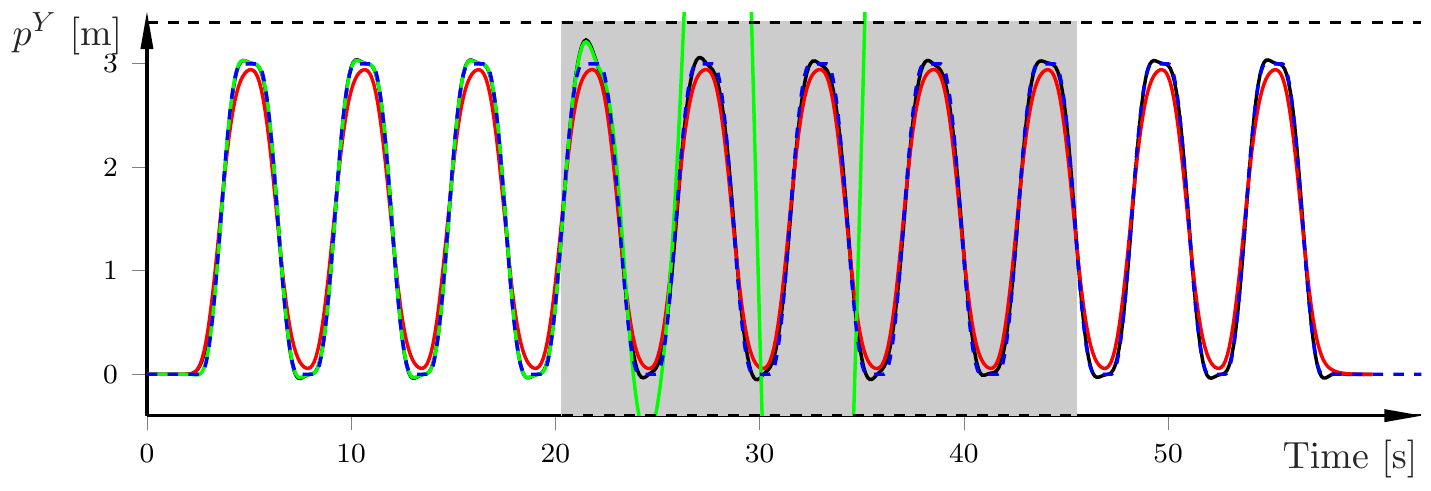}%
			{\tikzcustomremake{CarSimTraj}{\figdir\CarSimTraj}}
			{}%
			{\input{\figdir/CarSimTraj.tikz}}
		\end{flushright}
	\end{minipage}%

	\vspace*{-16pt}
	\caption{\red{CarSim results of lateral position for different controllers during the lane-change maneuvers. Black and blue dashed lines are the constraints and reference, respectively. The gray area indicates snow surface. The black, red, and green lines indicate the trajectories for \textsc{Friction-SNMPC}, \textsc{Snow-NMPC}, and \textsc{Asphalt-NMPC}.}}\label{fig:SMPCRes}
\end{figure}
\begin{figure}
	\input{tikzdefs}

	\input{tikztrim}
	\begin{minipage}{\textwidth} 
		\begin{flushright}
			
			\newcommand{\pgfigwidth}{0.85\textwidth}
			\newcommand{\pgfigheight}{9cm}
			\tikzsetnextfilename{cost}
			\filemodCmp{\figdir/cost.tikz}{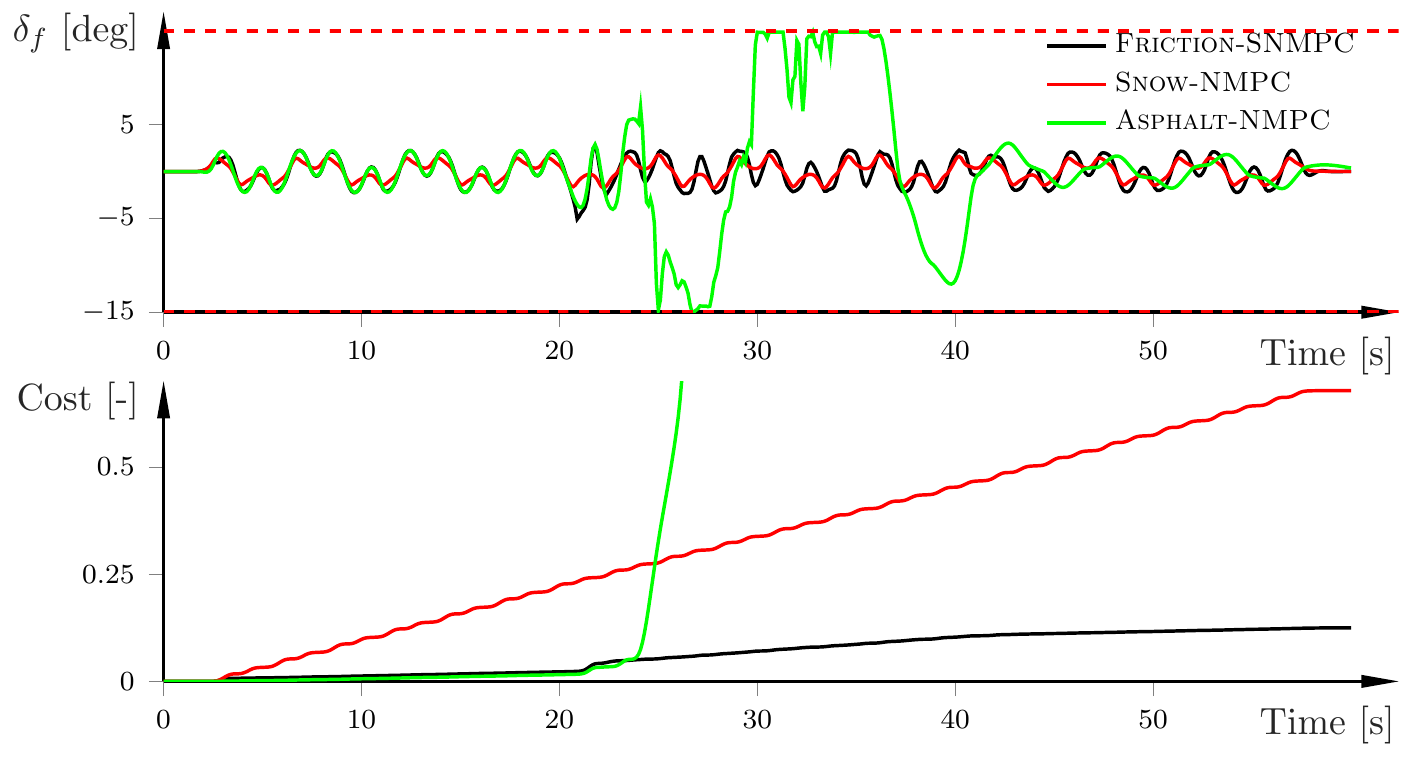}%
			{\tikzcustomremake{cost}{\figdir\cost}}
			{}%
			{\input{\figdir/cost.tikz}}
		\end{flushright}
	\end{minipage}%

	\vspace*{-15pt}
	\caption{\red{Front wheel steering angle commands and closed-loop cost for the CarSim validation results of the vehicle maneuver with surface changes as illustrated in Fig.~\ref{fig:SMPCRes}.}}\label{fig:cost}
	\vspace*{-10pt}
\end{figure}

\red{The performance of the estimator has been validated in several papers~\cite{Berntorp2020a,Berntorp2021d}. Here, we focus on the closed-loop control performance. Fig.~\ref{fig:SMPCRes} shows a comparison of the lateral position tracking performance between the three different controllers.
The controller denoted by \textsc{Asphalt-NMPC} shows significant overshoot and poor tracking performance once the surface switches to snow (indicated by the gray area).
The results demonstrate that the reference trajectory is tracked well by the \textsc{Friction-SNMPC} and \textsc{Snow-NMPC} controllers, and that there are some transients in \textsc{Friction-SNMPC} due to the convergence time of the tire-friction estimator.
The \textsc{Snow-NMPC} controller performs conservatively as expected. } 

\red{Fig.~\ref{fig:cost} displays the wheel steering angle and closed-loop cost for the three controllers. Note that the MPC cost function consists of more terms than the lateral position tracking error, and focusing solely on the lateral position can lead to wrong conclusions about performance. Fig.~\ref{fig:cost} shows that the surface change at about $21$s destabilizes the vehicle when using \textsc{Asphalt-NMPC}. Also, we can see that the adaptive \textsc{Friction-SNMPC} improves performance compared to the \textsc{Snow-NMPC} throughout the maneuver, reducing the closed-loop cost from over $0.6$ to about $0.1$.}

\subsection{Real-time Feasibility for Embedded Implementation}\label{ssec:RTtest}

To conclude the evaluation, we assess the real-time computational feasibility of the proposed methods. In 
Tables~\ref{tab:timing_results_GP}~and~\ref{tab:timing_results_stiffness}, we show the closed-loop computation times of the different friction-adaptive 
SNMPC on a dSPACE MicroAutoBox-II rapid prototyping unit. The table includes the computation time 
for the same maneuver as before for two different velocity references and two sets of combinations of 
the control horizon length~$N_{\mathrm{MPC}}$, time prediction length $T$, and the number of particles~$N_{\mathrm{PF}}$ that give good performance for the respective velocity reference. From these results, it is clear that the computation time for both the SNMPC and the 
estimators scale linearly and the 
total time remains well below the desired sampling time of $50$ms. 
\begin{table}
	\centering
		\caption{Timing results for Algorithm~\ref{alg:1} using the GP-based tire-friction estimator for two sets of combinations of 
		control horizon~$N_{\mathrm{MPC}}$, prediction length $T$, and number of particles~$N_{\mathrm{PF}}$ on a dSPACE 
		MicroAutoBox-II.}
	\label{tab:timing_results_GP}
	\begin{tabular}{lcccc}
		\toprule
	&	\multicolumn{2}{c}{$\begin{matrix}
				N_{\mathrm{MPC}}=15\\ T=0.75$s$ \\ N_{\mathrm{PF}}=200
			\end{matrix}$} & \multicolumn{2}{c}{$\begin{matrix}
			N_{\mathrm{MPC}}=20\\ T=1$s$ \\ N_{\mathrm{PF}}=100
		\end{matrix}$}\\ 
		& Mean & Max  	& Mean & Max \\
		\midrule
QP iterations				&	$3.65$	&	$25$ & $6.57$ &	$25$\\
 SNMPC	&	$16.4$ms	&	$29$ms	& $23.1$ms& $37.8$ms\\
Tire-friction estimator			&	$10.4$ms	&	$10.5$ms& $5.5$ms&	$5.6$ms\\
		\midrule
		Total turnaround time				&	$26.7$ms	&	$39.4$ms & $28.6$ms& $43.3$ms  \\
		\bottomrule
	\end{tabular}

\end{table}

\begin{table}
	\centering
	\caption{Timing results for Algorithm~\ref{alg:1} using the tire-stiffness estimator for two sets of combinations of 
		control horizon~$N_{\mathrm{MPC}}$, prediction length $T$, and number of particles~$N_{\mathrm{PF}}$ on a dSPACE 
		MicroAutoBox-II.}
	\label{tab:timing_results_stiffness}
	\begin{tabular}{lcccc}
		\toprule
		&	\multicolumn{2}{c}{$\begin{matrix}
				N_{\mathrm{MPC}}=15\\ T=0.75$s$ \\ N_{\mathrm{PF}}=1000
			\end{matrix}$} & \multicolumn{2}{c}{$\begin{matrix}
				N_{\mathrm{MPC}}=20\\ T=1$s$ \\ N_{\mathrm{PF}}=500
			\end{matrix}$}\\ 
		& Mean & Max  	& Mean & Max \\
		\midrule
		QP iterations				&	$4.03$	&	$20$ & $3.97$ &	$25$\\
		SNMPC	&	$10.4$ms	&	$20.6$ms	& $13$ms& $32.7$ms\\
		Tire-stiffness estimator			&	$12.2$ms	&	$14.1$ms& $6.2$ms&	$7$ms\\
		\midrule
		Total turnaround time				&	$22.6$ms	&	$34.2$ms & $19.1$ms& $39.5$ms   \\
		\bottomrule
	\end{tabular}

\end{table}

\section{Conclusion}
The proposed friction-adaptive control strategy uses a recently developed, computationally efficient SNMPC formulation. In this paper, the proposed SNMPC models the tire-road friction as an external disturbance and is flexible in the sense that it can be combined with any friction estimator, as long as the estimator is capable of outputting the first two moments of the friction estimate. We implemented the control strategy by leveraging two recently developed Bayesian friction estimators that model different aspects of the tire-friction characteristics. The first estimator is based on GPs and models the full nonlinear tire-friction curve, whereas the second estimator restricts the estimation of the tire stiffness. However, note that the proposed control strategy is not limited to these two estimators

Our Monte-Carlo studies show that friction-adaptive control based on SNMPC is generally superior to a nominal NMPC in terms of tracking error and constraint violations, and that the tire-friction estimator is superior to the tire-stiffness estimator. However, while the tire-friction estimator has more potential than the estimator based on a linear friction model, our results indicate that the GP-based tire-friction estimator should be combined with SNMPC and not nominal NMPC, while the tire-stiffness estimator seems to work sufficiently well irrespective of choosing SNMC or nominal NMPC. While these results alone are inconclusive, as they were obtained with a specific set of parameters and for a particular maneuver, they show that care should be taken with respect to parameter choices and the controller to integrate with, when using an estimator targeting the full tire-friction curve. 

The timing results, which we obtained using a dSPACE MicroAutoBox-II, indicate that the presented method is suitable for implementation on automotive embedded platforms. While the tire-friction estimator is more computationally demanding than the tire-stiffness estimator, both can run online with a sufficiently large number of particles. Furthermore, it is the SNMPC and not the estimator that takes up most of the computational demands. Hence, computational considerations do not seem to be the most important factor in determining which estimator to employ.
\label{sec:conclusion}
\bibliographystyle{tfnlm}
\bibliography{berntorpRef.bib}

\begin{thebibliography}{10}
\providecommand{\url}[1]{\normalfont{#1}}
\providecommand{\urlprefix}{Available from: }

\bibitem{Paden2016}
Paden~B, Cap~M, Yong~SZ, et~al. A survey of motion planning and control
  techniques for self-driving urban vehicles. {IEEE} Trans Intell Veh.
  2016;\hspace{0pt}1(1):33--55.

\bibitem{Berntorp2017c}
Berntorp~K. Path planning and integrated collision avoidance for autonomous
  vehicles. In: Amer. Control Conf.; May; Seattle, WA; 2017.

\bibitem{Lefevre2014}
Lef{\`e}vre~S, Vasquez~D, Laugier~C. A survey on motion prediction and risk
  assessment for intelligent vehicles. Robomech J. 2014;\hspace{0pt}1(1):1.

\bibitem{Okamoto2017a}
Okamoto~K, Berntorp~K, {Di Cairano}~S. Driver intention-based vehicle treat
  assessment using random forests and particle filtering. In: {IFAC} World
  Congress; Jul.; Toulouse, France; 2017.

\bibitem{Dicairano2018}
Di~Cairano~S, Kolmanovsky~IV. Real-time optimization and model predictive
  control for aerospace and automotive applications. In: Amer. Control Conf.;
  Jun.; Milwaukee, WI; 2018.

\bibitem{Hrovat2012}
Hrovat~D, Di~Cairano~S, Tseng~HE, et~al. The development of model predictive
  control in automotive industry: A survey. In: Int. Conf. Control
  Applications; Dubrovnik, Croatia; 2012.

\bibitem{Falcone2007}
Falcone~P, Borrelli~F, Asgari~J, et~al. Predictive active steering control for
  autonomous vehicle systems. IEEE Trans Control Syst Technol.
  2007;\hspace{0pt}15(3):566--580.

\bibitem{Berntorp2019f}
Berntorp~K, Quirynen~R, Uno~T, et~al. Trajectory tracking for autonomous
  vehicles on varying road surfaces by friction-adaptive nonlinear model
  predictive control. Veh Syst Dyn. 2019;\hspace{0pt}58(5):705--725.

\bibitem{Carvalho2015}
Carvalho~A, Lef{\'e}vre~S, Schildbach~G, et~al. Automated driving: The role of
  forecasts and uncertainty - a control perspective. Eur J Control.
  2015;\hspace{0pt}24:14--32.

\bibitem{DiCairano2016}
Di~Cairano~S, Kalabi{\'c}~U, Berntorp~K. Vehicle tracking control on
  piecewise-clothoidal trajectories by {MPC} with guaranteed error bounds. In:
  Conf. Decision and Control; Dec.; Las Vegas, NV; 2016.

\bibitem{Borrelli2005}
Borrelli~F, Falcone~P, Keviczky~T, et~al. {MPC}-based approach to active
  steering for autonomous vehicle systems. Int J Veh Auton Syst.
  2005;\hspace{0pt}3(2--4):265--291.

\bibitem{Berntorp2014}
Berntorp~K, Olofsson~B, Lundahl~K, et~al. Models and methodology for optimal
  trajectory generation in safety-critical road--vehicle manoeuvres. Veh Syst
  Dyn. 2014;\hspace{0pt}52(10):1304--1332.

\bibitem{Quirynen2018a}
Quirynen~R, Berntorp~K, {Di Cairano}~S. Embedded optimization algorithms for
  steering in autonomous vehicles based on nonlinear model predictive control.
  In: Amer. Control Conf.; Jun.; Milwaukee, WI; 2018.

\bibitem{Gustafsson1997}
Gustafsson~F. Slip-based tire-road friction estimation. Automatica.
  1997;\hspace{0pt}33:1087--1099.

\bibitem{Svendenius2007}
Svendenius~J. Tire modeling and friction estimation [dissertation]. Dept.
  Automatic Control, Lund University, Sweden; 2007.

\bibitem{Gustafsson2009}
Gustafsson~F. Automotive safety systems. IEEE Signal Processing magazine. 2009
  July;\hspace{0pt}26(4):32--47.

\bibitem{Rasmussen2006}
Rasmussen~C, Williams~C. Gaussian processes for machine learning. Cambridge,
  MA, USA: MIT Press; 2006.

\bibitem{Berntorp2018h}
Berntorp~K, {Di Cairano}~S. Tire-stiffness and vehicle-state estimation based
  on noise-adaptive particle filtering. IEEE Trans Control Syst Technol.
  2018;\hspace{0pt}27(3):1100--1114.

\bibitem{Berntorp2021d}
Berntorp~K. Online {B}ayesian inference and learning of {G}aussian-process
  state-space models. Automatica. 2021;\hspace{0pt}129:109613.

\bibitem{Doucet2009}
Doucet~A, Johansen~AM. A tutorial on particle filtering and smoothing: Fifteen
  years later. In: Crisan~D, Rozovsky~B, editors. Handbook of nonlinear
  filtering. Oxford University Press; 2009.

\bibitem{Quirynen2020}
Quirynen~R, Di~Cairano~S. {PRESAS}: Block-structured preconditioning of
  iterative solvers within a primal active-set method for fast model predictive
  control. Optimal Control Applications and Methods. 2020;\hspace{0pt}.

\bibitem{Feng2020}
Feng~X, {Di Cairano}~S, Quirynen~R. Inexact adjoint-based {SQP} algorithm for
  real-time stochastic nonlinear {MPC}. In: IFAC World Congress; Jul.; Berlin,
  Germany; 2020.

\bibitem{Vaskov2021}
Vaskov~S, Berntorp~K, Quirynen~R. Cornering stiffness adaptive, stochastic
  nonlinear model predictive control for vehicles. In: Amer. Control Conf.;
  May; New Orleans, LA; 2021.

\bibitem{Vaskov2022}
Vaskov~S, Quirynen~R, Menner~M, et~al. Friction-adaptive stochastic predictive
  control for trajectory tracking of autonomous vehicles. In: Amer. Control
  Conf.; Jun.; Atlanta, GA; 2022. Submitted.

\bibitem{Rajamani2006}
Rajamani~R. Vehicle dynamics and control. Springer-Verlag; 2006.

\bibitem{Berntorp2020a}
Berntorp~K. Online {B}ayesian tire-friction learning by {G}aussian-process
  state-space models. In: IFAC World Congress; Jul.; Berlin, Germany; 2020.

\bibitem{Solin2020}
Solin~A, S{\"a}rkk{\"a}~S. Hilbert space methods for reduced-rank {G}aussian
  process regression. Statistics and Computing.
  2020;\hspace{0pt}30(2):419--446.

\bibitem{Svensson2017}
Svensson~A, Sch{\"o}n~TB. A flexible state-space model for learning nonlinear
  dynamical systems. Automatica. 2017;\hspace{0pt}80:189--199.

\bibitem{Pacejka2006}
Pacejka~HB. Tire and vehicle dynamics. 2nd ed. Oxford, United Kingdom:
  Butterworth-Heinemann; 2006.

\bibitem{kerrigan2000soft}
Kerrigan~EC, Maciejowski~JM. Soft constraints and exact penalty functions in
  model predictive control. In: Control 2000 Conference, Cambridge; 2000. p.
  2319--2327.

\bibitem{Ahn2012}
Ahn~C, Peng~H, Tseng~HE. Robust estimation of road friction coefficient using
  lateral and longitudinal vehicle dynamics. Veh Syst Dyn.
  2012;\hspace{0pt}50(6):961--985.

\bibitem{Telen2015}
Telen~D, Vallerio~M, Cabianca~L, et~al. Approximate robust optimization of
  nonlinear systems under parametric uncertainty and process noise. J Proc
  Control. 2015;\hspace{0pt}33:140--154.

\bibitem{Andersson2018}
Andersson~JAE, Gillis~J, Horn~G, et~al. Casadi--a software framework for
  nonlinear optimization and optimal control. Mathematical Programming
  Computation. 2018;\hspace{0pt}.

\bibitem{Nocedal2006}
Nocedal~J, Wright~SJ. Numerical optimization. 2nd ed. Springer; 2006. Springer
  Series in Operations Research and Financial Engineering.

\bibitem{Diehl2005}
Diehl~M, Findeisen~R, Allg\"ower~F, et~al. Nominal stability of the real-time
  iteration scheme for nonlinear model predictive control. IEE Proc-Control
  Theory Appl. 2005;\hspace{0pt}152(3):296--308.

\bibitem{Gros2016}
Gros~S, Zanon~M, Quirynen~R, et~al. From linear to nonlinear {MPC}: bridging
  the gap via the real-time iteration. International Journal of Control.
  2020;\hspace{0pt}93(1):62--80.

\bibitem{Berntorp2018d}
Berntorp~K, Hoang~T, Quirynen~R, et~al. Control architecture design of
  autonomous vehicles. In: Conf. Control Technol. and Applications; Aug.;
  Copenhagen, Denmark; 2018. Invited paper.

\bibitem{Iso2002}
{ISO 3888-2: 2002}. Passenger cars--test track for a severe lane change
  manoeuvre--part 2: Obstacle avoidance ; 2002.

\bibitem{carsim}
{Mechanical Simulation}. Carsim [\url{www.carsim.com}]; 2021.

\bibitem{Berntorp2014a}
Berntorp~K. Particle filtering and optimal control for vehicles and robots
  [dissertation]. Department of Automatic Control, Lund University, Sweden;
  2014.

\end{thebibliography}

\end{document}